%% file: manuscript.tex
\documentclass[%
11pt,
reprint,
onecolumn,
tightenlines,
superscriptaddress,
preprintnumbers,
nofootinbib,
amsmath,amssymb,amsthm,
physrev,
eqsecnum,tikz,
]{revtex4-2}

\input{preamble}

\hypersetup{hidelinks}

\begin{document}

\preprint{To appear in Journal of the Mechanics and Physics of Solids (DOI: \href{https://doi.org/10.1016/j.jmps.2026.106909}{10.1016/j.jmps.2026.106909})}

\title{A Transport-Constrained Gradient-Flow Description of Continuum Dislocation Dynamics}

\author{Paria Behnoud \orcidlink{0009-0000-9597-9087}}
    \email{pbehnoud@cmu.edu}
    \affiliation{Program in Computational Mechanics, Carnegie Mellon University}
    \affiliation{Department of Civil and Environmental Engineering, Carnegie Mellon University}

\author{Anthony Rollett}
    \affiliation{Department of Materials Science and Engineering, Carnegie Mellon University}

\author{Carlos Mora-Corral}
    \affiliation{Departamento de Matem\'aticas, Universidad Aut\'onoma de Madrid}

\author{Kaushik Dayal \orcidlink{0000-0002-0516-3066}}
    \email{Kaushik.Dayal@cmu.edu}
    \affiliation{Department of Civil and Environmental Engineering, Carnegie Mellon University}
    \affiliation{Center for Nonlinear Analysis, Department of Mathematical Sciences, Carnegie Mellon University}
    \affiliation{Department of Mechanical Engineering, Carnegie Mellon University}

\date{\today}

\begin{abstract}
    Gradient flow frameworks are fundamental to numerous models of defect and microstructure evolution.
    This work develops a gradient flow formulation for continuum dislocation dynamics, with the key distinction that the admissible rates are transport-constrained.
    Using the dislocation density and compatible slip fields as the primary state variables, their admissible rates are restricted by geometric conservation laws. 
    The corresponding transport velocities are then selected variationally through a constrained Onsager principle.
    The resulting model has both a variational gradient flow structure and consistency with the geometric dislocation conservation statement; existing models of dislocation evolution typically have either one or the other feature.
    
    This thermodynamic structure provides a dislocation driving force that includes elastic stress contributions -- corresponding to the classical Peach--Koehler force -- as well as contributions from the core and nonconvex lattice energies.
    A simplified version of the model is used to study the energetics and dynamics of {an opposite-sign dislocation pair. The lattice energy keeps the cores compact, and an external applied shear either assists annihilation or, above a threshold, separates the pair.}
\end{abstract}

\maketitle

\section{Introduction}
\label{sec:introduction}

Dislocation evolution is a central problem in the mechanics of crystalline solids. 
The motion, interaction, and collective organization of dislocations
govern plastic deformation, strain hardening, size effects, and the emergence of microstructure. 
A fundamental goal is therefore to formulate dislocation dynamics from principles of energy, dissipation, and kinematics, e.g., \cite{mcdowell2024nonequilibrium-1,Koslowski2002}.

Variational principles, such as gradient flows, have played a central role in continuum theories of materials evolution, e.g., \cite{Carter1997,Mielke2008,Fife1990}, and established this viewpoint as a powerful
framework for describing dissipative evolution in materials.
For example, in models of grain-boundary motion
\cite{KinderlehrerGrain2001,KinderlehrerMesoscale2004}, the evolution law is
often obtained by combining a free energy with an appropriate dissipation
mechanism or metric structure. More generally, Onsager-type variational
principles provide a systematic way to combine thermodynamic driving forces
with dissipation mechanisms~\cite{Onsager1931,Masao2012}.
In such approaches, the choice of free energy, together with the
gradient-flow metric, dissipation mechanism, and admissible class of
variations, determines the dissipative evolution.

Dislocation mechanics presents an important and interesting setting for this framework. 
The elastic energy depends on the elastic distortion, while the defect content is encoded by the incompatibility of the plastic distortion. 
Thus the energetic driving force is tied to both elastic equilibrium and the geometry of dislocations. 
However, the evolution of dislocations cannot be described by arbitrary changes of the plastic distortion: dislocations move by transport, and their admissible evolution must respect the conservation and compatibility structure associated with the dislocation density tensor \cite{mura1963continuous}.

This provides the motivation for the present work: to formulate a variational principle in which the motion is driven by energy and dissipation, but the admissible directions are constrained by dislocation transport.
The central question is how to construct a gradient-flow-type theory for dislocations in which the velocity of transport, rather than an arbitrary rate of plastic distortion, is selected by a variational principle.

\subsection{Prior Approaches to Dislocation Evolution}

Variational perspectives and related thermodynamic approaches have played a key role in the formulation of models of dislocation dynamics, and plasticity broadly.
An early work was by Rickman and Vi\~nals~\cite{RickmanVinals1997}, who developed a related coarse-grained dissipative model in which the dislocation-density tensor evolves according to Burgers-vector conservation and a phenomenological flux chosen to decrease the plastic free energy.
Koslowski, Cuiti\~no, and Ortiz~\cite{Koslowski2002} developed a phase-field model with gradient-flow dynamics for dislocation evolution, incorporating long-range elastic interactions and core effects.
Gurtin~\cite{Gurtin2004} developed a gradient theory of small-deformation plasticity based on virtual power and microforce balance, in which the free energy depends on the elastic response and on the Burgers tensor, or equivalently on the curl of the plastic distortion. 
Cermelli and Gurtin \cite{Cermelli1999TheMO} used maximal dissipation to deduce the dynamics of dislocations, which was then analyzed mathematically by Blass et al.\ \cite{blass2014} and Alicandro et al.\ \cite{Alicandro2016DynamicsOD}.
More recently, Dondl, Kurzke, and Wojtowytsch~\cite{Dondl2018} studied the gradient flow of a phase-field model for plastic slip in the presence of forest dislocations and showed that the limiting dynamics may differ from the gradient flow of the limiting energy, with the obstacles generating an effective frictional resistance. 

Continuum dislocation dynamics has also emphasized the transport character of dislocation evolution. 
The transport viewpoint underlying continuum dislocation dynamics has its origins in the conservation laws for Burgers-vector density and dislocation flux introduced by Kosevich~\cite{Kosevich1962} and the transport law for the dislocation density tensor introduced by Mura \cite{mura1963continuous,mura2013micromechanics}. 
These seminal kinematic structures provide the starting point for many subsequent continuum descriptions of dislocation evolution.
Modern continuum dislocation dynamics formulations further emphasize the role of transport, closure assumptions, and mobility laws in connecting dislocation density evolution to crystal plasticity~\cite{hochrainer2014continuum,El-Azab2020}.

Many continuum theories build on this transport structure but close the model by prescribing the velocity or current through an additional constitutive kinetic law, with some key references cited below.
Groma and Balogh~\cite{Groma1999} derived transport equations for positive and negative edge dislocation densities starting from overdamped equations of motion for discrete dislocations. 
Limkumnerd and Sethna developed a continuum theory in which the Nye dislocation density tensor evolves through a topological current derived from Peach--Koehler-driven glide, and demonstrated the formation of sharp dislocation walls~\cite{Sethna2006,Sethna2007,Sethna2008}.
Chen et al.~\cite{Yong2013} developed a three-dimensional scaling theory of continuum dislocation dynamics leading to self-organized fractal pattern formation. 
Hochrainer, Zaiser, and Gumbsch~\cite{Hochrainer2007} introduced higher-dimensional continuum dislocation dynamics to represent curved dislocation lines, with velocities determined by Peach--Koehler forces, mobility laws, and possible line-tension terms. 
Monavari and Zaiser~\cite{Monavari2018} formulated a slip-system-specific form of the dislocation transport law in which
the glide direction is fixed kinematically, while the scalar velocity is
supplied constitutively. 
Sandfeld and Zaiser~\cite{Sandfeld2015} developed a
minimal continuum dislocation plasticity model involving total density,
geometrically necessary dislocation density, and curvature density, closed by a
thresholded mobility law. 
Xia and El-Azab~\cite{Xia2015} used an explicit
evolution law based on the dislocation transport law and prescribed the scalar speed through a linear
mobility relation involving the resolved shear stress. 
More recently, Zhang,
Wu, and Zaiser~\cite{Yufan2025} reformulated continuum dislocation dynamics as
a phase-field theory with conserved order parameters and applied it to
dislocation patterning.
A related class of approaches represents individual dislocation lines
through evolving level-set functions: e.g., Chen et al.\
\cite{2010_Chen} developed a three-dimensional level-set formulation in
which the dislocation motion is overdamped and the velocity is prescribed
through a mobility tensor with distinct glide and climb components acting
on a Peach--Koehler force. The total driving stress includes self-stress,
applied stress, and stresses generated by misfitting obstacles.

At the level of state variables, continuum descriptions have
also emphasized the role of geometrically necessary dislocations in
size-dependent plasticity.
Gao and Huang~\cite{GaoHuang2003} and Arsenlis and
Parks~\cite{arsenlis1999crystallographic,arsenlis2002modeling}
developed formulations in which geometrically necessary dislocation content
plays an important role in size-dependent plasticity and crystal-plasticity
response. This provides a related use of dislocation density as a measure of
incompatibility. 
A complementary approach is due to Rosakis~\cite{Rosakis2001}, who analyzed dislocations in an augmented continuum model and developed kinetic relations between applied shear stress and dislocation speed, providing a comparison at the level of force--velocity closure.

\subsection{Contributions of This Work}

The distinguishing feature of the present formulation is that we develop a gradient flow structure for dislocation dynamics that also respects the geometric conservation laws of dislocation transport.
Specifically, the dislocation transport law defines the admissible tangent directions of the variational evolution, and the
associated transport velocities are selected through a constrained Onsager principle.
While there are dislocation evolution models that have a gradient flow structure, they typically do not obey the geometric conservation law of dislocation dynamics; conversely, dislocation dynamics models that are based on the geometric conservation law do not have a gradient-flow structure.

We develop a formulation in which the primary independent state variables are the dislocation density field \(\bfalpha\) and the compatible slip field \(\mathbold R\), with the plastic distortion $\bfbeta_p$ being constructed from $\bfalpha$ and $\bfR$.
The admissible rates of the theory are induced by transport laws: $\bfalpha$ is transported as a vector-valued two-form, while \(\mathbold R\) is transported row-wise as a vector-valued one-form. 
{Since the transport law for $\bfalpha$ determines the plastic-distortion rate only up to a gradient, Mura's rate~\cite{mura1963continuous} corresponds to setting this gradient to zero; here, it is given instead by the transport of $\mathbold R$, and Mura's choice is recovered when $\mathbold R$ is not transported.} 
We use a constrained Onsager principle to select the transport velocities variationally, with the variations taken along the transport-admissible directions.
Minimization of the Onsager functional determines the optimal velocities associated with \(\bfalpha\) and \(\mathbold R\), and thereby yields a transport-constrained gradient-flow system. 

The distinguishing feature is the organization of the evolution as a gradient flow on the transport-admissible tangent space, with a state-dependent metric determined by the dissipation due to the transport velocities. 
The Onsager construction identifies a state-dependent metric on the transport-admissible tangent space and interprets the evolution as steepest descent of the reduced free energy with respect to this metric. 
The variation of the total reduced free energy is restricted to the tangent space generated by transport satisfying the dislocation transport law, and the dissipation potential defines a state-dependent metric on that tangent
space. 
In essence, our formulation replaces the usual thermodynamic force--velocity viewpoint with an explicit metric structure for the admissible evolution.
This additional structure provides a natural basis for variational time discretizations and for future analysis of the metric and well-posedness properties of the evolution.

The resulting driving force for \(\bfalpha\) has the structure of a generalized Peach--Koehler force, with additional thermodynamic contributions generated by the defect (core) energy and nonconvex lattice energy. 
In the limit in which these additional energetic contributions vanish, the formulation recovers the classical Peach--Koehler force for a single dislocation line. 
The model therefore embeds the standard force law of dislocation mechanics within the transport-constrained variational framework, with the classical Peach--Koehler closure recovered as a special energetic limit.

\begin{remark}[Wasserstein gradient flows and their Riemannian formulation]
    The present construction is closely related in spirit to Wasserstein gradient flows, where dissipative evolution is formulated as the steepest descent of a free energy subject to a transport constraint. 
    The scheme introduced by Jordan, Kinderlehrer, and Otto showed that certain dissipative partial differential equations can be interpreted as gradient flows of free energy with respect to the Wasserstein metric~\cite{JKO1998}. 
    Otto subsequently developed the formal Riemannian interpretation of such transport gradient
    flows, in which the continuity equation induces a metric structure on
    admissible tangent directions~\cite{Otto2000}. Later developments clarified
    the analytical and variational structure of Wasserstein gradient flows and their connection to transport, continuity equations, and
    variational time discretizations
    \cite{AmbrosioGigliSavare2008,Santambrogio2015,Santambrogio2017,
    Friesecke2024}.
    
    The analogy with the present formulation lies in the role of transport in
    defining admissible tangent directions. In the Wasserstein setting, density
    rates are generated through a continuity equation and the associated
    transport velocity. Here, admissible changes of the dislocation density
    are generated by dislocation velocities through the dislocation transport law,
    while the compatible slip is evolved through its own
    transport law. The corresponding metric therefore measures the dissipation
    required to realize transport-admissible variations rather than arbitrary
    variations.
\end{remark}

\subsection{Structure of the Paper}

\begin{itemize}
    \item
    Section~\ref{sec:formulation} lays out the kinematics, specifically the relation of the plastic distortion $\bfbeta_p$ to the dislocation density tensor \(\bfalpha\) and the compatible slip \(\mathbold R\), and the free energy for dislocation plasticity.

    \item
    Section~\ref{sec:lie_derivative_defect_variables} develops the Lie-derivative interpretation of the defect variables and derives the transport-admissible tangent directions for \(\bfalpha\) and \(\mathbold R\).

    \item
    Section~\ref{sec:constrained_gradient_flow_dislocation_evolution} formulates the transport-constrained Onsager principle, and derives the corresponding driving forces and the resulting transport equations.
    It also identifies the associated transport manifold, metric, formal gradient-flow structure, and energy-dissipation identity.

    \item
    Section \ref{sec:PK} compares the derived driving force on $\bfalpha$ with the classical Peach--Koehler force.
    The roles of the lattice energy and the core energy are discussed, and the driving force is shown to reduce to the classical expression in the case of a single straight dislocation when the lattice energy is neglected.
    
    \item 
    Section \ref{sec:comparison_cahn_hilliard} compares the structure of the current with the classical Cahn--Hilliard approach in phase-field modeling of conserved species.

    \item
    Section~\ref{sec:examples} {applies the model to the annihilation of two opposite-sign dislocations on a single glide plane.
    The pair glides toward annihilation, and a superposed uniform resolved shear stress either assists annihilation or, above a threshold, separates the pair.
    }
    
\end{itemize}

\section{Kinematics and Free Energy Formulation}
\label{sec:formulation}

\subsection{Notation}
\label{sec:notation}

We work in the geometrically linear setting.
We use \(\delta\) to denote a variation, while \(D\mathcal F[\mathbold z][\mathbold s]\) denotes the directional derivative
of a functional \(\mathcal F\) at a generic state \(\mathbold z\) in the direction \(\mathbold s\).
We use the Einstein convention that repeated indices are summed unless otherwise stated.
{
Throughout the paper, we assume that all functions have sufficient regularity for all operations being performed to be well defined.
}

\subsection{Kinematics}
\label{sec:kinematics}

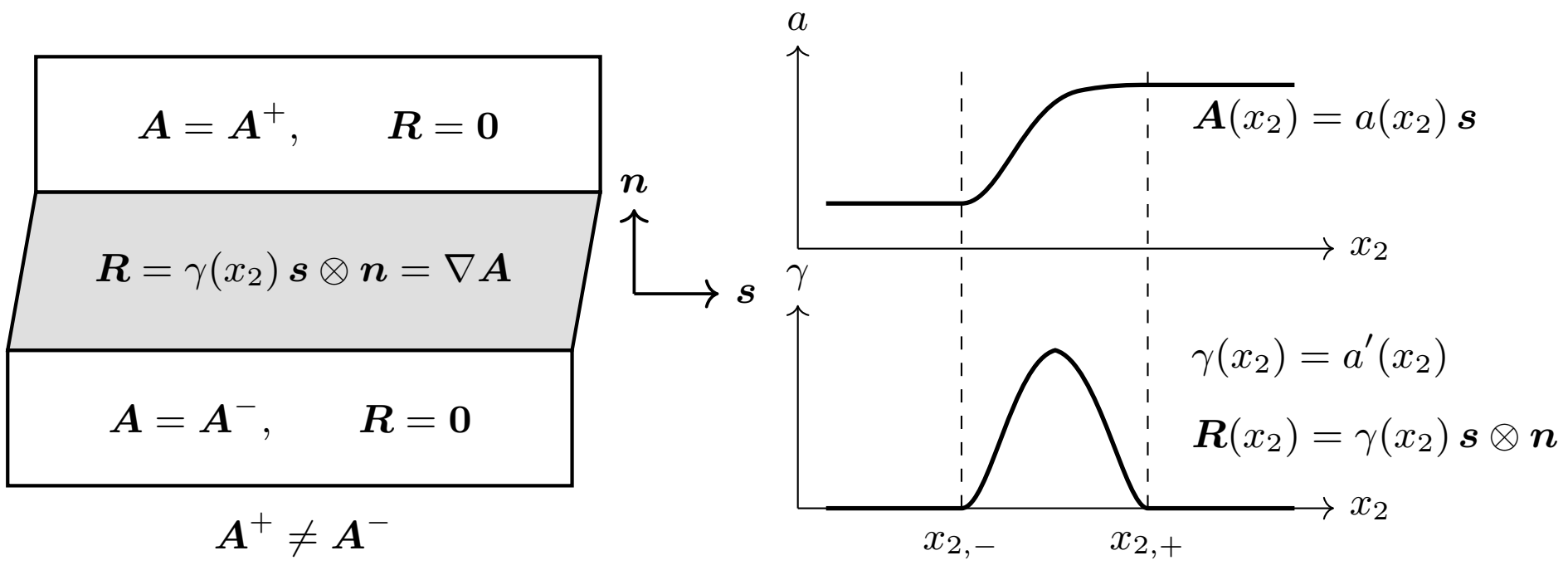
\begin{figure}[htb!]
\centering
\begin{tikzpicture}[scale=0.90, every node/.style={font=\small}]

\draw[line width=0.8pt]
    (0.45,2.95) rectangle (5.45,4.15);

\fill[gray!25]
    (0.45,2.95) --
    (5.45,2.95) --
    (5.20,1.55) --
    (0.20,1.55) -- cycle;

\draw[line width=0.8pt]
    (0.45,2.95) --
    (5.45,2.95) --
    (5.20,1.55) --
    (0.20,1.55) -- cycle;

\draw[line width=0.8pt]
    (0.20,0.35) rectangle (5.20,1.55);

\node at (2.95,3.55)
{
    $\mathbold A=\mathbold A^{+},
    \qquad
    \mathbold R=\bfzero$
};

\node at (2.70,0.95)
{
    $\mathbold A=\mathbold A^{-},
    \qquad
    \mathbold R=\bfzero$
};

\node at (2.82,2.25)
{
    $\mathbold R
    =
    \gamma(x_2)\,\mathbold s\otimes\mathbold n
    =
    \nabla\mathbold A$
};

\coordinate (orient) at (5.75,2.05);

\draw[->,line width=0.7pt]
    (orient) -- ++(0.75,0)
    node[right] {$\mathbold s$};

\draw[->,line width=0.7pt]
    (orient) -- ++(0,0.75)
    node[above] {$\mathbold n$};

\node at (2.82,-0.08)
{
    $\mathbold A^{+}\neq\mathbold A^{-}$
};

\node at (2.82,-0.62)
{
    \textbf{(a)}
};

\begin{scope}[xshift=7.2cm]

\draw[->]
    (0,2.45) -- (4.75,2.45)
    node[right] {$x_2$};

\draw[->]
    (0,2.45) -- (0,4.25)
    node[above] {$a$};

\draw[dashed]
    (1.45,0.15) -- (1.45,4.05);

\draw[dashed]
    (3.10,0.15) -- (3.10,4.05);

\draw[line width=1.0pt]
    (0.25,2.85) -- (1.45,2.85)
    .. controls (1.80,2.85) and (2.00,3.75) ..
    (2.50,3.85)
    .. controls (2.75,3.90) and (2.95,3.90) ..
    (3.10,3.90)
    -- (4.40,3.90);

\node[anchor=west] at (3.35,3.60)
{
    $\mathbold A(x_2)=a(x_2)\,\mathbold s$
};

\draw[->]
    (0,0.15) -- (4.75,0.15)
    node[right] {$x_2$};

\draw[->]
    (0,0.15) -- (0,1.95)
    node[above] {$\gamma$};

\draw[line width=1.0pt]
    (0.25,0.15) -- (1.45,0.15)
    .. controls (1.70,0.15) and (1.90,1.45) ..
    (2.28,1.55)
    .. controls (2.66,1.45) and (2.90,0.15) ..
    (3.10,0.15)
    -- (4.40,0.15);

\node[anchor=west] at (3.35,1.50)
{
    $\gamma(x_2)=a'(x_2)$
};

\node[anchor=west] at (3.35,0.78)
{
    $\mathbold R(x_2)
    =
    \gamma(x_2)\,\mathbold s\otimes\mathbold n$
};

\node at (1.45,-0.18)
{
    $x_{2,-}$
};

\node at (3.10,-0.18)
{
    $x_{2,+}$
};

\node at (2.35,-0.62)
{
    \textbf{(b)}
};

\end{scope}

\end{tikzpicture}

    \caption{
        Schematic example of a compatible slip through simple shear, showing plastic slip and nonzero $\bfbeta_p$ without the presence of dislocations.
        The compatible part of the plastic distortion is represented by
        $\mathbold R=\nabla\mathbold A$.
        In (a), $\mathbold A$ is constant outside the shaded layer and varies within it.
        As an illustrative example, we take $\mathbold A(x_2)=a(x_2)\,\mathbold s$, where $\mathbold s$ denotes the slip direction and $\mathbold n$ the layer normal.
        It follows that $\mathbold R=\gamma(x_2)\,\mathbold s\otimes\mathbold n$, with $\gamma(x_2)=a'(x_2)$, so that $\mathbold R$ vanishes outside the layer and is localized within it.
        In (b), we see the corresponding one-dimensional profiles.
    }
\label{fig:compatible_layer}
\end{figure}

{
The body $\Omega$ is a bounded, smooth, simply connected\footnote{
    {The case of dislocations in non-simply-connected domains is studied in \cite{YavariGoriely2012,SozioYavari2023}.}
} domain with connected boundary $\partial\Omega$ and outward unit normal $\mathbold n$; curl, divergence, gradient, and cross products act row-wise.
The state variables are the displacement $\mathbold u$, the dislocation density tensor $\bfalpha$, and the compatible slip $\mathbold R$, subject to
\begin{equation}
    \divergence\bfalpha=\bfzero,
    \qquad
    \curl\mathbold R=\bfzero
    \quad\text{in }\Omega.
    \label{eq:state_constraints}
\end{equation}
Since $\Omega$ is simply connected, $\mathbold R=\nabla\mathbold A$ for a field $\mathbold A$ determined up to a constant~\cite{Yavari2013}.

Following the Stokes--Helmholtz decomposition in dislocation mechanics~\cite{Kroner1958,kroner2010continuum,Wenzelburger1998,BinzSchwarzWenzelburger2001}, we define the plastic distortion
\begin{equation}
    \bfbeta_p
    :=
    \mathcal C[\bfalpha]
    +
    \mathbold R,
    \quad\text{where}\quad
    \mathcal C[\bfalpha](\mathbold x)
    :=
    \curl\int_\Omega\mathbold G(\mathbold x,\mathbold y)\,\bfalpha(\mathbold y) \dm\Omega_{\mathbold y}
    \label{eq:beta_p_reconstruction_notation}
\end{equation}
and $\mathbold G$ is the Green's function of the vector Laplacian in $\Omega$, applied to each row of $\bfalpha$, with the boundary conditions below.
That is, $\mathbold H:=\int_\Omega\mathbold G\,\bfalpha\,\mathrm d\Omega_{\mathbold y}$ solves
\begin{equation}
    -\Delta\mathbold H=\bfalpha
    \quad\text{in }\Omega,
    \qquad
    \mathbold H\times\mathbold n=\bfzero,
    \quad
    \divergence\mathbold H=\bfzero
    \quad\text{on }\partial\Omega,
    \label{eq:green_problem_C}
\end{equation}
which has exactly one solution because $\partial\Omega$ is connected~\cite{Amrouche1998}.
By \eqref{eq:state_constraints}, $\divergence\mathbold H$ is harmonic in $\Omega$ and vanishes on $\partial\Omega$, so $\divergence\mathbold H=\bfzero$ and $\curl\curl\mathbold H=-\Delta\mathbold H=\bfalpha$.
Hence $\mathcal C[\bfalpha]=\curl\mathbold H$ satisfies
\begin{equation}
    \curl\mathcal C[\bfalpha]=\bfalpha,
    \quad
    \divergence\mathcal C[\bfalpha]=\bfzero
    \quad\text{in }\Omega,
    \qquad
    \mathcal C[\bfalpha]\mathbold n=\bfzero
    \quad\text{on }\partial\Omega,
    \label{eq:div_curl_problem_C}
\end{equation}
where the boundary condition follows from $\mathbold H\times\mathbold n=\bfzero$.
The solution of this div--curl problem is unique because $\Omega$ is simply connected~\cite{Amrouche1998}.
From \eqref{eq:state_constraints}--\eqref{eq:div_curl_problem_C},
\begin{equation}
    \curl\bfbeta_p=\bfalpha,
    \label{eq:alpha_definition}
\end{equation}
so $\bfalpha$ is the dislocation density of $\bfbeta_p$.
The field $\mathcal C[\bfalpha]$ is the incompatible part of $\bfbeta_p$, determined nonlocally by $\bfalpha$, and $\mathbold R$ is its compatible part.
Conversely, every $\bfbeta_p$ has exactly one such decomposition, with $\bfalpha=\curl\bfbeta_p$ and $\mathbold R=\bfbeta_p-\mathcal C[\curl\bfbeta_p]$.

The displacement gradient decomposes additively as
\begin{equation}
    \nabla \mathbold u
    =
    \bfbeta_e
    +
    \bfbeta_p ,
    \label{eq:additive_distortion_decomposition}
\end{equation}
which defines the elastic distortion $\bfbeta_e$.

Figure~\ref{fig:compatible_layer} schematically illustrates a simple-shear example of the compatible part $\mathbold R=\nabla\mathbold A$.
}

\subsubsection{Boundary Conditions}
\label{subsubsec:boundary_conditions}

The boundary $\partial\Omega$ consists of disjoint parts $\partial\Omega_D$ and $\partial\Omega_N$, on which the displacement $\bar{\mathbold u}$ and the traction $\bar{\mathbold t}$ are prescribed:
\begin{equation}
\mathbold u=\bar{\mathbold u}
\quad\text{on }\partial\Omega_D,
\qquad
\bfsigma^*\mathbold n=\bar{\mathbold t}
\quad\text{on }\partial\Omega_N ,
\label{eq:mechanical_boundary_conditions}
\end{equation}
where $\bfsigma^*$ is the stress \eqref{eq:stress_definition_transport_formulation}.
If $\partial\Omega_D$ is empty, the applied tractions have zero net force and moment.
On $\partial\Omega$, we further impose, for the transport velocities $\mathbold V_{\bfalpha}$ and $\mathbold V_R$ of Section~\ref{sec:lie_derivative_defect_variables},
\begin{equation}
\mathbold R\mathbold V_R=\bfzero,
\qquad
\mathbold P_R^{\mathrm d}\mathbold n=\bfzero,
\label{eq:boundary_conditions_on_boundary}
\end{equation}
and either $(\bfalpha\times\mathbold V_{\bfalpha})\times\mathbold n=\bfzero$ or $\mathbold P_\alpha^{\mathrm d}\times\mathbold n=\bfzero$ (microscopically hard or free~\cite{Gurtin2004}), where $\mathbold P_\alpha^{\mathrm d}$ and $\mathbold P_R^{\mathrm d}$ are defined in \eqref{eq:defect_and_lattice_microstresses_alpha_R}.
In Section~\ref{sec:comparison_cahn_hilliard}, $\mathbold J\cdot\mathbold n=0$ on $\partial\Omega$.
Under these conditions, all boundary terms arising from integration by parts vanish. Surface energies, such as those of the steps left where dislocations exit the body, are not included.

\subsection{Energetics}
\label{sec:energetics}

We use the general free energy:
\begin{equation}
    \mathcal F[\mathbold u,\bfalpha,\mathbold R]
    :=
    \int_{\Omega} W_e\left(\nabla\mathbold u-\bfbeta_p\right)\,\mathrm d\Omega
    +
    \int_{\Omega} W_{\alpha}\left(\bfalpha\right)\,\mathrm d\Omega
    +
    \int_{\Omega} W_R\left(\nabla\mathbold R\right)\,\mathrm d\Omega
    +
    \int_{\Omega} \Psi\left(\bfbeta_p\right)\,\mathrm d\Omega
    -
    \int_{\partial\Omega_N} \bar{\mathbold t}\cdot\mathbold u\,\mathrm d\Gamma ,
    \label{eq:free_energy_alpha_R}
\end{equation}
where the plastic distortion is reconstructed according to \eqref{eq:beta_p_reconstruction_notation}, \(\bfbeta_p=\mathcal C[\bfalpha]+\mathbold R\). {Here, the terms $W_e$, $W_\alpha$, $W_R$, and $\Psi$ are stored-energy densities; dissipation is represented by the velocity-dependent terms in the Onsager functional \eqref{eq:onsager_functional_alpha_R}.}

For simplicity, the elastic energy density $W_e (\bfbeta_e)$ is taken to have the quadratic form:
\begin{equation}
    W_e
    \left(
    \nabla\mathbold u-\bfbeta_p
    \right)
    :=
    \frac{1}{2}
    \left(
    \nabla\mathbold u-\bfbeta_p
    \right)
    :
    \mathbold C
    \left(
    \nabla\mathbold u-\bfbeta_p
    \right),
    \label{eq:elastic_energy_density}
\end{equation}
where \(\mathbold C\) is the fourth-order elasticity tensor.
The corresponding stress is then defined as:
\begin{equation}
    \bfsigma
    :=
    \mathbold C \bfbeta_e
    =
    \mathbold C \left(\nabla\mathbold u-\bfbeta_p\right) 
    =    
    \mathbold C \left(\nabla\mathbold u-\mathcal C[\bfalpha]-\mathbold R\right).
    \label{eq:stress_definition}
\end{equation}

The defect-energy density \(W_\alpha(\bfalpha)\) is a function of the dislocation density and represents the dislocation core energy.
Core regularization and dislocation-density-dependent energetic effects have
entered continuum dislocation theories through several distinct routes. 
In nonsingular continuum models of dislocations, the dislocation core is spatially regularized, removing the classical elastic singularity while retaining a Peach--Koehler-type force evaluated from the regularized stress field \cite{Cai2006,lazar2005nonsingular,lazar2013fundamentals, po2018non,Agiasofitou2017}.
Defect or core energy contributions have been naturally introduced through free energy densities that are functions of the dislocation density tensor, e.g. \cite{Gurtin2004}.

The lattice-energy density $\Psi(\bfbeta_p)$ is a nonconvex
function representing crystallographic slip resistance
relative to a prescribed reference
lattice~\cite{clayton2011nonlinear,EpsteinElzanowski2007}.
We assume $\Psi(\bfbeta_p+\mathbold S)=\Psi(\bfbeta_p)$ for
the prescribed lattice-equivalent slip increments
$\mathbold S$~\cite{Ericksen1980,PitteriZanzotto2003};
under a change of global reference configuration,
the lattice symmetries must be transformed consistently.
Similar Peierls-type nonconvex lattice energies
have been used to represent crystallographic slip in several prior models
\cite{Koslowski2002,WangJinCuitinoKhachaturyan2001,RodneyLeBouarFinel2003,ShenWang2004,Hunter2011,Peng2020,Rosakis2001,Rosakis2013}.

Finally, \(W_R(\nabla\mathbold R)\) is an energy associated with spatial gradients of the compatible slip $\bfR$.
For standard models of dislocation plasticity, \textit{this contribution vanishes}, i.e., $W_R \equiv 0$.
However, we retain it in the model for generality as it can be used to represent phase transformations when these are coupled to dislocation plasticity.

Free energies containing these, or closely related, energetic elements have appeared in continuum, gradient, and phase-field descriptions of dislocations. 
In particular, elastic energies coupled to the Burgers tensor or dislocation density have been considered in continuum dislocation plasticity models, e.g., \cite{Gurtin2004,GromaCsikorZaiser2003,Zaiser2015,Hochrainer2016}.

\begin{remark}[Contribution of the Compatible Slip to the Energy]
    The compatible slip contributes to the plastic distortion \(\bfbeta_p=\mathcal C[\bfalpha]+\mathbold R\), and hence to the free energy through the elastic energy \(W_e\) and the lattice energy \(\Psi\), regardless of whether the additional regularization term \(W_R\) is retained.
\end{remark}

For later use, we define the energy-conjugate quantities:
\begin{equation}
    \mathbold P_\alpha^{\mathrm d} := \frac{\partial W_\alpha}{\partial\bfalpha},
    \qquad
    \mathbold P_R^d := \frac{\partial W_R}{\partial\nabla\mathbold R},
    \qquad
    \mathbold P_\Psi := \parderiv{\Psi}{\bfbeta_p}.
    \label{eq:defect_and_lattice_microstresses_alpha_R}
\end{equation}

\subsubsection{Reduced Energy by Minimizing out the Displacement}

For fixed \((\bfalpha,\mathbold R)\), the displacement field \(\mathbold u^*\) is determined by minimizing the free energy with respect to \(\mathbold u\).
This reflects the physical statement that we expect mechanical equilibrium to be achieved much faster than the scale of defect relaxation. 
Therefore, we are essentially at mechanical equilibrium\footnote{
    {
    In regimes where inertia is accounted for, the equation of momentum balance does not follow from a minimum principle.
    However, it is possible -- in principle -- to eliminate the displacement using momentum balance, and then apply a minimum principle for defect evolution.
    }
} at all times as the defect density evolves.

The equilibrium displacement $\bfu^*[\bfalpha,\bfR]$ is obtained as the solution of:
\begin{subequations}
\begin{align}
    \divergence\bfsigma^*
    =
    \bfzero,
    & \text{ in } \Omega,
    \\
    \bfsigma^*\mathbold n
    =
    \bar{\mathbold t},
    & \text{ on } \partial\Omega_N,
     \\
    {\mathbold u^*
    =
    \bar{\mathbold u},}
    & {\text{ on } \partial\Omega_D,}
\end{align}
\label{eq:elastic_equilibrium_transport_formulation}
\end{subequations}
where:
\begin{equation}
    \bfsigma^*
    :=
    \left.\frac{\partial W_e}{\partial\nabla\bfu}\right|_{\bfu^*}.
    \label{eq:stress_definition_transport_formulation}
\end{equation}

Eliminating the displacement field through elastic equilibrium defines the
reduced free energy
\begin{equation}
\mathcal F^{\mathrm{red}}
[
\bfalpha,\mathbold R
]
:=
\mathcal F
[
\mathbold u^*,
\bfalpha,
\mathbold R
].
\label{eq:reduced_free_energy_alpha_R}
\end{equation}

\section{Defect Transport Laws through Lie Derivatives}
\label{sec:lie_derivative_defect_variables}

We now describe the geometric origin of the transport constraints: we write transport laws for \(\bfalpha\) and \(\mathbold R\), and the admissible rate for $\bfbeta_p$ is induced using \eqref{eq:beta_p_reconstruction_notation}.
{We use a Lie-derivative approach that provides a common framework for transporting both $\bfalpha$ and the compatible slip $\mathbold R$. 
It also connects naturally to the transport structure of Wasserstein gradient flows.}

\subsection{Lie Derivative for Dislocation Density}
\label{subsec:lie_derivative_alpha}

Following Hochrainer et al \cite{Hochrainer2007}, we outline the dislocation transport law for $\alpha$ using Lie derivatives.
The dislocation density tensor is a vector-valued two-form that measures Burgers-vector content through oriented
surface elements:
\begin{equation}
    \bfalpha
    =
    \frac{1}{2}
    \alpha^a{}_{ij}
    \,
    \mathbold e_a\otimes
    \mathrm dx^i\wedge \mathrm dx^j .
    \label{eq:alpha_vector_valued_two_form}
\end{equation}
For a differential form \(\bfomega\), Cartan's formula gives
\begin{equation}
\mathcal L_{\mathbold V}\bfomega
=
\mathrm d
\left(
\iota_{\mathbold V}\bfomega
\right)
+
\iota_{\mathbold V}
\left(
\mathrm d\bfomega
\right),
\label{eq:cartan_formula_general}
\end{equation}
where \(\iota_{\mathbold V}\) denotes contraction with the vector field
\(\mathbold V\). {Applying this to \(\bfalpha\), transported by the dislocation velocity \(\mathbold v\), gives
\begin{equation}
\mathcal L_{\mathbold v}
\bfalpha
=
\mathrm d
\left(
\iota_{\mathbold v}
\bfalpha
\right)
+
\iota_{\mathbold v}
\left(
\mathrm d\bfalpha
\right).
\label{eq:cartan_formula_alpha}
\end{equation}
}
Since
\(\bfalpha=\mathrm d\bfbeta_p\), we have
\begin{equation}
    \mathrm d\bfalpha
    =
    \mathrm d^2\bfbeta_p
    =
    \bfzero 
    \iff
    \divergence\bfalpha
    =
    \bfzero,
    \label{eq:d_alpha_zero}
\end{equation}
expressing the fact that dislocation lines cannot end inside the body.

Under the standard identification $\alpha^a{}_{ij}=\epsilon_{ijk}\alpha^a_k$, with $\epsilon_{ijk}$ the permutation symbol, the contraction is
\begin{equation}
\iota_{\mathbold v}
\bfalpha
=
\bfalpha
\times
\mathbold v,
\label{eq:interior_product_alpha_cross_velocity}
\end{equation}
where the cross product is row-wise: $\left(\bfalpha\times\mathbold v\right)_i=\bfalpha_i\times\mathbold v$.
We write the transport in terms of $\mathbold V_{\bfalpha}:=-\mathbold v$, the negative of the dislocation velocity, which gives the form of the transport law used throughout this paper. 
Thus
\begin{equation}
\mathcal L_{\mathbold v}
\bfalpha
=
\curl
\left(
\bfalpha
\times
\mathbold v
\right)
=
-
\curl
\left(
\bfalpha
\times
\mathbold V_{\bfalpha}
\right).
\label{eq:lie_derivative_alpha_curl_flux}
\end{equation}

The Lie-transport equation provides the classical dislocation transport law of Kosevich and Mura~\cite{Kosevich1962,mura1963continuous,mura2013micromechanics}:
\begin{equation}
    \frac{\partial\bfalpha}{\partial t}
    +
    {\mathcal L_{\mathbold v}}
    \bfalpha
    =
    \bfzero
    \iff
    \dot{\bfalpha}
    =
    \curl
    \left(
    \bfalpha
    \times
    \mathbold V_{\bfalpha}
    \right).
\label{eq:alpha_tangent_direction}
\end{equation}
This gives the admissible tangent direction for \(\bfalpha\).
We define the associated dislocation flux as
$\mathbold J_\alpha:=\bfalpha\times\mathbold V_{\bfalpha}$,
so that $\dot{\bfalpha}=\curl\mathbold J_\alpha$.

\subsection{Transport of the Compatible Slip}
\label{subsec:transport_R}

The geometric viewpoint used here is motivated by transport-based formulations of microstructural evolution, in which the relevant geometric object is evolved through an appropriate transport or conservation law; see, for example, the interface-density balance formulation of Agrawal and Dayal~\cite{AgrawalDayal2015PartI,AgrawalDayal2015PartII}.
In their setting, the gradient of a phase field is evolved through a conservation statement for interfaces. 
Here, we adopt the same transport-based viewpoint for \(\mathbold R\), which is evolved directly as a state variable.

Accordingly, we transport \(\mathbold R\) row-wise as a vector-valued
one-form. Let \(\mathbold V_R\) denote the velocity field associated with the
transport of \(\mathbold R\); in Figure \ref{fig:compatible_layer}, this would correspond to the vertical motion of the slip surfaces as interfaces\footnote{
    Under an applied shear, it is possible -- though typically unlikely -- that the shear is accommodated through the coherent interfacial motion of slip surfaces, but is much more likely it is accommodated by the compatible slip left behind by the motion of dislocations. To suppress the former deformation mechanism, we can set $\bfV_R\equiv \bfzero$ or equivalently that the corresponding mobility coefficient vanishes.
}.

Since \(\mathbold R=\nabla\mathbold A\), each row of
\(\mathbold R\) is a closed one-form. Hence, by Cartan's formula,
\begin{equation}
\mathcal L_{\mathbold V_R}\mathbold R
=
\mathrm d
\left(
\iota_{\mathbold V_R}\mathbold R
\right).
\label{eq:lie_derivative_R}
\end{equation}
In components,
\begin{equation}
\left(
\mathcal L_{\mathbold V_R}\mathbold R
\right)_{ij}
=
V_R^k\partial_k R_{ij}
+
R_{ik}\partial_j V_R^k
=
\partial_j
\left(
R_{ik}V_R^k
\right),
\label{eq:lie_derivative_R_components}
\end{equation}
where the last equality uses
\(\partial_kR_{ij}=\partial_jR_{ik}\), which follows from
\(\curl\mathbold R=\bfzero\).
{Dislocation motion also generates compatible slip, e.g., the slip left behind a gliding dislocation. The evolution equation for the compatible slip is therefore}
\begin{equation}
\frac{\partial \mathbold R}{\partial t}
+
\mathcal L_{\mathbold V_R}\mathbold R
=
{\mathcal P[\bfalpha\times\bfV_{\bfalpha}]},
\label{eq:R_lie_transport}
\end{equation}
or equivalently,
\begin{equation}
\dot R_{ij}
=
{\bigl(\mathcal P[\bfalpha\times\bfV_{\bfalpha}]\bigr)_{ij}}
-
\partial_j
\left(
R_{ik}V_R^k
\right),
\label{eq:R_tangent_direction}
\end{equation}
{where, for a field $\mathbold X$, $\mathcal P[\mathbold X]:=\nabla\mathbold w$, and $\mathbold w$, determined up to a constant, solves row-wise
\begin{equation}
\Delta\mathbold w=\divergence\mathbold X
\quad\text{in }\Omega,
\qquad
(\nabla\mathbold w)\mathbold n=\mathbold X\mathbold n
\quad\text{on }\partial\Omega.
\label{eq:P_definition}
\end{equation}
Section~\ref{subsec:induced_rate_beta_p} shows that $\mathcal P[\mathbold X]$ is the compatible part of $\mathbold X$ and derives the source $\mathcal P[\bfalpha\times\bfV_{\bfalpha}]$.}
Thus \(\mathbold R\) is transported by \(\mathbold V_R\) {and generated by dislocation motion}, with the condition \(\curl\mathbold R=\bfzero\) preserved by the evolution{, since both terms on the right of \eqref{eq:R_tangent_direction} are gradients}.
This structure is analogous to the gradient-based interface-density balance used by Agrawal and Dayal~\cite{AgrawalDayal2015PartI}.

\subsection{Induced Rate of the Plastic Distortion}
\label{subsec:induced_rate_beta_p}

The transport equations for \(\bfalpha\) and \(\mathbold R\) induce an
admissible rate for the plastic distortion. Using
\eqref{eq:beta_p_reconstruction_notation}, differentiation gives
\begin{equation}
\dot{\bfbeta}_p
=
\mathcal C[\dot{\bfalpha}]
+
\dot{\mathbold R}.
\label{eq:beta_p_rate_from_reconstruction}
\end{equation}
Thus the rate of the plastic distortion has an incompatible part induced by the
transport of \(\bfalpha\) and a compatible part {carried
by \(\mathbold R\)}.

Using the tangent direction induced by the dislocation transport law \eqref{eq:alpha_tangent_direction}, the first term in \eqref{eq:beta_p_rate_from_reconstruction} is a lift of \(\dot{\bfalpha}\) from dislocation-density space to plastic-distortion space.
{Since $\mathcal C$ is linear and independent of time, \eqref{eq:beta_p_reconstruction_notation} and \eqref{eq:alpha_tangent_direction} give
\begin{equation}
\begin{aligned}
\mathcal C[\dot{\bfalpha}](\mathbold x)
&=
\curl\int_\Omega\mathbold G(\mathbold x,\mathbold y)\,\dot{\bfalpha}(\mathbold y)\dm\Omega_{\mathbold y}
=
\curl\int_\Omega\mathbold G(\mathbold x,\mathbold y)\,\curl\bigl(\bfalpha\times\bfV_{\bfalpha}\bigr)(\mathbold y)\dm\Omega_{\mathbold y}
\\
&=
\bfalpha\times\bfV_{\bfalpha}
-
\mathcal P[\bfalpha\times\bfV_{\bfalpha}],
\end{aligned}
\label{eq:C_alpha_dot}
\end{equation}
where $\mathcal P$ is defined in \eqref{eq:P_definition}.
For any $\mathbold X$, the field $\mathbold X-\mathcal P[\mathbold X]$ has curl $\curl\mathbold X$, zero divergence, and zero normal component on $\partial\Omega$, so it is the unique solution of the div--curl problem \eqref{eq:div_curl_problem_C} with $\bfalpha$ replaced by $\curl\mathbold X$. Hence $\mathcal C[\curl\mathbold X]=\mathbold X-\mathcal P[\mathbold X]$, which gives the last equality in \eqref{eq:C_alpha_dot} for $\mathbold X=\bfalpha\times\bfV_{\bfalpha}$, and $\mathcal P[\mathbold X]$ is the compatible part of $\mathbold X$.
The term $\mathcal P[\bfalpha\times\bfV_{\bfalpha}]$ is thus the compatible slip generated by dislocation motion; it enters the rate of $\mathbold R$ as the source in \eqref{eq:R_lie_transport}.}
Substituting \eqref{eq:C_alpha_dot} and \eqref{eq:R_lie_transport} into \eqref{eq:beta_p_rate_from_reconstruction}, the generated terms cancel, and the induced admissible plastic-distortion rate is
\begin{equation}
\bfeta
:=
\dot{\bfbeta}_p
=
\bfalpha \times \bfV_{\bfalpha}
-
\nabla
\left(
\mathbold R\bfV_R
\right).
\label{eq:eta_alpha_R_lie}
\end{equation}
Thus the admissible plastic-distortion rate is determined by the two transport velocities \((\bfV_{\bfalpha},\bfV_R)\).
This is the rate-level tangent direction used in the constrained Onsager principle below. 

{Since the curl of a gradient vanishes, the transport law \eqref{eq:alpha_tangent_direction} determines $\dot\bfbeta_p$ only up to a gradient. The first term in \eqref{eq:eta_alpha_R_lie}, which includes the compatible slip generated by dislocation motion, is the plastic-distortion rate of Kosevich and Mura~\cite{Kosevich1962,mura1963continuous}, for which this gradient is zero; the second term is the gradient given by the transport of \(\mathbold R\). 
Physically, the former corresponds to plastic distortion evolution due to dislocation transport, and the latter corresponds to the plastic distortion evolution due to coherent interfacial motion of slip planes (if this occurs).
We also refer to Sozio and Yavari \cite{SozioYavari2023} for an interesting discussion related to the fixing of the gauge.}

\section{Constrained Gradient Flow for Dislocation Evolution}
\label{sec:constrained_gradient_flow_dislocation_evolution}

An unconstrained gradient flow permits arbitrary rates of the plastic distortion \(\bfbeta_p\). 
In contrast, the formulation developed
here restricts the evolution to the transport-generated directions derived in
Section~\ref{sec:lie_derivative_defect_variables}. The transport velocities,
rather than arbitrary rates of the state variables, are therefore the
kinematic variables over which the variational principle is posed.

The guiding analogy is the Wasserstein formulation of dissipative evolution,
in which admissible state rates are generated by transport velocities and the
velocity is selected variationally through an Onsager
principle~\cite{JKO1998,AmbrosioGigliSavare2008,Santambrogio2015,
Onsager1931,Masao2012}. In the present setting, the ordinary continuity
equation is replaced by the transport laws for \(\bfalpha\) and
\(\mathbold R\). The transport velocities therefore generate the admissible
tangent directions and serve as the variational unknowns. We first derive the
resulting transport-constrained Onsager evolution and subsequently identify
its associated metric structure analogous to the Wasserstein Riemannian formulation.

\subsection{Transport-Constrained Onsager Principle}
\label{subsec:transport_constrained_onsager_principle}

We now apply Onsager's variational principle~\cite{Onsager1931,Masao2012} to
the transport-admissible velocities. The state variables are
\((\bfalpha,\mathbold R)\), and the plastic distortion is reconstructed
according to \eqref{eq:beta_p_reconstruction_notation}. Thus,
\(\bfbeta_p\) is not varied independently; its rate is induced by the
admissible rates of \(\bfalpha\) and \(\mathbold R\) following \eqref{eq:eta_alpha_R_lie}.

We begin by defining
\begin{equation}
    s_{\bfalpha}
    :=
    \curl
    \left(
    \bfalpha\times\mathbold V_{\bfalpha}
    \right),
    \qquad
    \left(s_R\right)_{ij}
    :=
    {\left(\mathcal P[\bfalpha\times\mathbold V_{\bfalpha}]\right)_{ij}}
    -
    \partial_j
    \left(
    R_{ik}V_R^k
    \right)
    \label{eq:transport_rates_for_onsager}
\end{equation}
as the transport-admissible rates generated by the velocities
\((\mathbold V_{\bfalpha},\mathbold V_R)\). 
Then, the first variation of the reduced free energy along an admissible tangent
direction \((s_{\bfalpha},s_R)\) is
\begin{equation}
    D\mathcal F^{\mathrm{red}}[\bfalpha,\mathbold R][s_{\bfalpha},s_R]
    =
    \int_\Omega\left[
        \left(-\bfsigma^*+\mathbold P_\Psi\right):\bfeta 
        +
        \mathbold P_\alpha^{\mathrm d}:s_{\bfalpha}
        +
        \mathbold P_R^{\mathrm d}:\nabla s_R
    \right]
    \dm\Omega .
\label{eq:first_variation_reduced_energy_alpha_R}
\end{equation}
For brevity, we define the effective thermodynamic stress-like quantity
\begin{equation}
    \bfsigma^{\mathrm{eff}}
    :=
    \bfsigma^*
    -
    \mathbold P_\Psi .
    \label{eq:effective_stress_with_lattice_energy}
\end{equation}

The Onsager functional is therefore
\begin{equation}
\mathcal J
[
\mathbold V_{\bfalpha},\mathbold V_R
]
:=
\frac{1}{2}
\int_{\Omega}
\mathbold V_{\bfalpha}
\cdot
\mathbold M_{\bfalpha}^{-1}
\mathbold V_{\bfalpha}
\,\mathrm d\Omega
+
\frac{1}{2}
\int_{\Omega}
\mathbold V_R
\cdot
\mathbold M_R^{-1}
\mathbold V_R
\,\mathrm d\Omega
+
D\mathcal F^{\mathrm{red}}
[
\bfalpha,\mathbold R
]
[
s_{\bfalpha},s_R
],
\label{eq:onsager_functional_alpha_R}
\end{equation}
with \(s_{\bfalpha}\) and \(s_R\) given by
\eqref{eq:transport_rates_for_onsager}, and with \(\bfeta\) given by \eqref{eq:eta_alpha_R_lie}. 
The tensors $\bfM_{\bfalpha}$ and $\mathbold M_R$ are the defect mobilities, which may depend on the state.
We assume decoupled kinetics for simplicity. More general coupled kinetics can be described by a symmetric positive-definite mobility operator $\mathbb M$ acting on $\mathbold V=(\mathbold V_{\bfalpha},\mathbold V_R)$. This gives $\mathbold V^*=\mathbb M\mathbold f$, where $\mathbold f=(\mathbold f_{\bfalpha},\mathbold f_R)$, and preserves the energy-dissipation structure.

The transport velocities are selected by minimizing this functional:
\begin{equation}
\left(
\mathbold V_{\bfalpha}^*,
\mathbold V_R^*
\right)
:=
\operatorname*{arg\,min}_{
\mathbold V_{\bfalpha},
\mathbold V_R
}
\mathcal J
[
\mathbold V_{\bfalpha},\mathbold V_R
].
\label{eq:optimal_velocities_definition_alpha_R}
\end{equation}
The resulting evolution is then obtained by substituting the minimizing
velocities into the transport equations:
\begin{equation}
\dot{\bfalpha}
=
\curl
\left(
\bfalpha\times\mathbold V_{\bfalpha}^*
\right),
\qquad
\dot R_{ij}
=
{\left(\mathcal P[\bfalpha\times\mathbold V_{\bfalpha}^*]\right)_{ij}}
-
\partial_j
\left(
R_{ik}(V_R^*)^k
\right) .
\label{eq:closed_transport_evolution_alpha_R}
\end{equation}
Thus, Onsager's principle selects the transport velocities of the constrained gradient flow rather than the arbitrariness of the rate of the plastic distortion.

\subsection{Driving Forces and Transport Closure}
\label{subsec:driving_forces_transport_closure}

We first derive the thermodynamic driving forces associated with the dislocation density and the compatible slip. 
We then use these forces in the Onsager principle to obtain the transport closure. 

\subsubsection{Driving Force for the Evolution of the Dislocation Density}
\label{subsubsec:alpha_driving_force}

We denote the reduced elastic energy contribution by
\begin{equation}
E^{\mathrm{el}}[\bfalpha,\mathbold R]
:=
\int_{\Omega}
W_e\left(\nabla\mathbold u^*-\bfbeta_p\right)
\,\mathrm d\Omega
-
\int_{\partial\Omega_N}
\bar{\mathbold t}\cdot\mathbold u^*
\,\mathrm d\Gamma ,
\label{eq:elastic_energy_definition_new}
\end{equation}
where
\(\bfbeta_p=\mathcal C[\bfalpha]+\mathbold R\) and
\(\mathbold u^*=\mathbold u^*[\bfalpha,\mathbold R]\) is the equilibrium displacement.

We likewise denote the lattice-energy contribution by
\begin{equation}
E^{\mathrm{lat}}[\bfalpha,\mathbold R]
:=
\int_{\Omega}
\Psi(\bfbeta_p)
\,\mathrm d\Omega .
\label{eq:lattice_energy_definition_new}
\end{equation}
We also denote the defect-energy contributions associated with
\(\bfalpha\) and \(\mathbold R\) by
\begin{equation}
E^{\alpha}[\bfalpha]
:=
\int_{\Omega}
W_{\alpha}(\bfalpha)
\,\mathrm d\Omega ,
\qquad
E^{R}[\mathbold R]
:=
\int_{\Omega}
W_R(\nabla\mathbold R)
\,\mathrm d\Omega .
\label{eq:defect_energy_definitions_new}
\end{equation}

Because \(\mathbold u^*\) satisfies elastic equilibrium, the variation of
the reduced elastic energy contribution along an admissible
plastic-distortion rate \(\bfeta\) is
\begin{equation}
\delta E^{\mathrm{el}}
=
-
\int_{\Omega}
\bfsigma^*:\bfeta
\,\mathrm d\Omega .
\label{eq:elastic_energy_rate_beta_p_new}
\end{equation}

The corresponding variation of the lattice energy is
\begin{equation}
\delta E^{\mathrm{lat}}
=
\int_{\Omega}
\parderiv{\Psi}{\bfbeta_p}:\bfeta
\,\mathrm d\Omega .
\label{eq:lattice_energy_rate_beta_p_new}
\end{equation}

Using \(\bfsigma^{\mathrm{eff}}\) defined in \eqref{eq:effective_stress_with_lattice_energy} and the induced plastic-distortion rate from \eqref{eq:eta_alpha_R_lie}, we have:
\begin{equation}
\delta E^{\mathrm{el}}+\delta E^{\mathrm{lat}}
=
-
\int_{\Omega}
\bfsigma^{\mathrm{eff}}
:
\left(
\bfalpha\times\mathbold V_{\bfalpha}
\right)
\,\mathrm d\Omega
+
\int_{\Omega}
\bfsigma^{\mathrm{eff}}
:
\nabla
\left(
\mathbold R\mathbold V_R
\right)
\,\mathrm d\Omega .
\label{eq:elastic_lattice_energy_rate_expanded_new}
\end{equation}

The defect-energy variation associated with \(\bfalpha\) is
\begin{equation}
\delta E^{\alpha}
=
\int_{\Omega}
\mathbold P_{\alpha}^{\mathrm d}
:
\curl
\left(
\bfalpha\times\mathbold V_{\bfalpha}
\right)
\,\mathrm d\Omega ,
\label{eq:alpha_defect_variation_new}
\end{equation}
where we have used \(\mathbold P_{\alpha}^{\mathrm d}\) defined in \eqref{eq:defect_and_lattice_microstresses_alpha_R}.
After integrating by parts under the boundary conditions of Section~\ref{subsubsec:boundary_conditions}, this can be written as
\begin{equation}
\delta E^{\alpha}
=
\int_{\Omega}
\mathbold V_{\bfalpha}
\cdot
\sum_i
\left[
\left(
\curl\mathbold P_{\alpha,i}^{\mathrm d}
\right)
\times
\bfalpha_i
\right]
\,\mathrm d\Omega .
\label{eq:alpha_defect_variation_force_form_new}
\end{equation}

Combining all of the variational terms gives the \(\bfalpha\)-driving force{; the term involving $\mathbold P_R^{\mathrm d}$ arises from $W_R$ through the compatible slip generated by dislocation motion, and is derived after \eqref{eq:div_PR_component_definition}}:
\begin{equation}
\mathbold f_{\bfalpha}
=
\sum_i
\left[
\left(
\bfsigma_i^{\mathrm{eff}}
-
\curl
\mathbold P_{\alpha,i}^{\mathrm d}
{
+
\bigl(\mathcal P[\divergence\mathbold P_R^{\mathrm d}]\bigr)_i
}
\right)
\times
\bfalpha_i
\right].
\label{eq:alpha_force_general_new}
\end{equation}

\subsubsection{Driving Force for the Evolution of the Compatible Slip}
\label{subsubsec:R_driving_force}

{For the compatible slip, we first consider the part of the admissible tangent direction generated by $\mathbold V_R$; the part generated by dislocation motion is treated after \eqref{eq:div_PR_component_definition}:}
\begin{equation}
\left(s_R\right)_{ij}
=
-
\partial_j
\left(
R_{ik}V_R^k
\right).
\label{eq:R_tangent_for_force_derivation_new}
\end{equation}

The combined elastic and lattice-energy contribution associated with this
tangent direction is
\begin{equation}
-
\int_{\Omega}
\bfsigma^{\mathrm{eff}}
:
s_R
\,\mathrm d\Omega
=
\int_{\Omega}
\sigma_{ij}^{\mathrm{eff}}
\partial_j
\left(
R_{ik}V_R^k
\right)
\,\mathrm d\Omega .
\label{eq:R_elastic_lattice_variation_new}
\end{equation}

The compatible slip energy contribution is
\begin{equation}
\delta E^{R}
=
\int_{\Omega}
\mathbold P_R^{\mathrm d}
:
\nabla s_R
\,\mathrm d\Omega ,
\label{eq:R_defect_variation_before_parts_new}
\end{equation}
where we use \(\mathbold P_R^{\mathrm d}\) defined in \eqref{eq:defect_and_lattice_microstresses_alpha_R}.
After integrating by parts {under the boundary conditions of Section~\ref{subsubsec:boundary_conditions}},
\begin{equation}
\delta E^{R}
=
-
\int_{\Omega}
\divergence\mathbold P_R^{\mathrm d}
:
s_R
\,\mathrm d\Omega
=
\int_{\Omega}
\left(
\divergence\mathbold P_R^{\mathrm d}
\right)_{ij}
\partial_j
\left(
R_{ik}V_R^k
\right)
\,\mathrm d\Omega .
\label{eq:R_defect_variation_after_parts_new}
\end{equation}

Therefore, the combined contribution associated with the compatible slip is
\begin{equation}
\int_{\Omega}
\left(
\bfsigma^{\mathrm{eff}}
+
\divergence\mathbold P_R^{\mathrm d}
\right)_{ij}
\partial_j
\left(
R_{ik}V_R^k
\right)
\,\mathrm d\Omega
=
-
\int_{\Omega}
\partial_j
\left(
\bfsigma^{\mathrm{eff}}
+
\divergence\mathbold P_R^{\mathrm d}
\right)_{ij}
R_{ik}V_R^k
\,\mathrm d\Omega .
\label{eq:R_variation_combined_new}
\end{equation}

This can be written as
\begin{equation}
-
\int_{\Omega}
\mathbold V_R
\cdot
\mathbold f_R
\,\mathrm d\Omega ,
\label{eq:R_variation_force_form_new}
\end{equation}
where
\begin{equation}
\left(
\mathbold f_R
\right)_k
=
R_{ik}
\partial_j
\left(
\bfsigma^{\mathrm{eff}}
+
\divergence\mathbold P_R^{\mathrm d}
\right)_{ij}.
\label{eq:R_force_general_new}
\end{equation}
Here, \(\bfsigma_i^{\mathrm{eff}}\),
\(\mathbold P_{\alpha,i}^{\mathrm d}\), and \(\bfalpha_i\) denote the
\(i\)-th rows of the corresponding tensor fields, and repeated indices are
summed. In addition,
\begin{equation}
\left(
\divergence\mathbold P_R^{\mathrm d}
\right)_{ij}
:=
\partial_k
\left(
\mathbold P_R^{\mathrm d}
\right)_{ijk}.
\label{eq:div_PR_component_definition}
\end{equation}
The part $\mathcal P[\bfalpha\times\mathbold V_{\bfalpha}]$ of $s_R$ generated by dislocation motion does not change $\bfeta$ and enters only through $W_R$. Integrating by parts as in \eqref{eq:R_defect_variation_after_parts_new}, and using that, by \eqref{eq:P_definition}, $\mathcal P$ is the $L^2$-orthogonal projection onto gradient fields, its contribution is
\begin{equation}
\int_\Omega\mathbold P_R^{\mathrm d}:\nabla\mathcal P[\bfalpha\times\mathbold V_{\bfalpha}]\,\mathrm d\Omega
=
-\int_\Omega\mathcal P[\divergence\mathbold P_R^{\mathrm d}]:(\bfalpha\times\mathbold V_{\bfalpha})\,\mathrm d\Omega
=
-\int_\Omega\mathbold V_{\bfalpha}\cdot\sum_i\bigl(\mathcal P[\divergence\mathbold P_R^{\mathrm d}]\bigr)_i\times\bfalpha_i\,\mathrm d\Omega ,
\label{eq:P_term_contribution}
\end{equation}
which gives the last term in \eqref{eq:alpha_force_general_new}. This term vanishes for $W_R=0$, as in Sections~\ref{sec:PK}--\ref{sec:examples}.

\subsubsection{Transport Closure from the Onsager Principle}
\label{subsubsec:onsager_transport_closure}

The driving forces in \eqref{eq:alpha_force_general_new} and \eqref{eq:R_force_general_new} have related precedents in the literature. The former has the generalized Peach--Koehler structure found in continuum dislocation mechanics~\cite{mura2013micromechanics}, with the defect-energy contribution having the same overall physical origina as the energetic back-stresses~\cite{Gurtin2004,Hochrainer2016}, while its lattice-energy contribution is related to Peierls-type models~\cite{Koslowski2002}. 
Since $\mathbold R=\nabla\bfA$, the transport part of its evolution corresponds to advection of $\bfA$, up to an additive time-dependent constant. 
Viewing $\mathcal F^{\mathrm{red}}[\bfalpha,\nabla\bfA]$ as a functional of $\bfA$ at fixed $\bfalpha$,
with variations of $\bfA$ that vanish near $\partial\Omega$, we can write
\eqref{eq:R_force_general_new} as
\begin{equation}
(\mathbold f_R)_k
=
R_{ik}\frac{\delta\mathcal F^{\mathrm{red}}}{\delta A_i},
\qquad
\frac{\delta\mathcal F^{\mathrm{red}}}{\delta A_i}
=
\partial_j
\left(
\bfsigma^{\mathrm{eff}}
+
\divergence\mathbold P_R^{\mathrm d}
\right)_{ij}.
\label{eq:fR_variational_derivative_A}
\end{equation}

The transport of the compatible gradient field
is kinematically analogous to interface-density
transport in Agrawal and Dayal
\cite{AgrawalDayal2015PartI}.
However, their interface kinetics are prescribed
constitutively, whereas here the transport velocity
is selected through the constrained Onsager principle.
Since \eqref{eq:R_lie_transport} makes
\eqref{eq:eta_alpha_R_lie} exact, both driving forces
are obtained from the exact reduced-energy
variation along the
transport-admissible rates.

Combining all energetic contributions {under the boundary conditions of Section~\ref{subsubsec:boundary_conditions}}, the
reduced-energy variation can be written as
\begin{equation}
\delta\mathcal F^{\mathrm{red}}
=
-
\int_{\Omega}
\mathbold V_{\bfalpha}
\cdot
\mathbold f_{\bfalpha}
\,\mathrm d\Omega
-
\int_{\Omega}
\mathbold V_R
\cdot
\mathbold f_R
\,\mathrm d\Omega .
\label{eq:reduced_energy_rate_force_form_new}
\end{equation}

Substitution into \eqref{eq:onsager_functional_alpha_R} gives
\begin{equation}
\mathcal J
=
\frac{1}{2}
\int_{\Omega}
\mathbold V_{\bfalpha}
\cdot
\mathbold M_{\bfalpha}^{-1}
\mathbold V_{\bfalpha}
\,\mathrm d\Omega
+
\frac{1}{2}
\int_{\Omega}
\mathbold V_R
\cdot
\mathbold M_R^{-1}
\mathbold V_R
\,\mathrm d\Omega
-
\int_{\Omega}
\mathbold V_{\bfalpha}
\cdot
\mathbold f_{\bfalpha}
\,\mathrm d\Omega
-
\int_{\Omega}
\mathbold V_R
\cdot
\mathbold f_R
\,\mathrm d\Omega .
\label{eq:onsager_functional_force_form_new}
\end{equation}
Stationarity with respect to the two transport velocities yields
\begin{equation}
\mathbold V_{\bfalpha}^*
=
\mathbold M_{\bfalpha}
\mathbold f_{\bfalpha},
\qquad
\mathbold V_R^*
=
\mathbold M_R
\mathbold f_R .
\label{eq:optimal_velocities_force_law_new}
\end{equation}

Substituting the optimal velocities into the transport laws gives the
transport-constrained evolution system
\begin{subequations}
\begin{align}
    \dot{\bfalpha}
    & =
    \curl\left(\bfalpha\times\mathbold M_{\bfalpha}\mathbold f_{\bfalpha}\right),
    \\
    \dot R_{ij}
    & =
    {\left(\mathcal P[\bfalpha\times\mathbold M_{\bfalpha}\mathbold f_{\bfalpha}]\right)_{ij}}
    -\partial_j\left[R_{ik}\left(\mathbold M_R\mathbold f_R\right)^k\right].
\end{align}
\label{eq:closed_evolution_alpha_R_new}
\end{subequations}
This system is the transport-constrained evolution generated by the Onsager principle. Its corresponding metric and gradient-flow interpretation,
analogous to the Riemannian formulation, are developed in Section~\ref{subsec:transport_manifold_metric}.

\subsection{Transport Manifold, Dissipation Metric, and Gradient-Flow Structure}
\label{subsec:transport_manifold_metric}

Having derived the transport-constrained Onsager evolution and its driving
forces, we now identify the associated geometric structure. The construction
below should be understood as a formal interpretation of the
resulting evolution rather than as a necessary element in its derivation.

Motivated by the formal Riemannian interpretation of Wasserstein gradient
flows~\cite{Otto2000}, we define a state-dependent transport metric for the variables
\((\bfalpha,\mathbold R)\). The role played by the continuity equation in the
Wasserstein setting is played here by the transport laws for \(\bfalpha\) and
the compatible slip \(\mathbold R\). The metric is therefore not imposed
directly on arbitrary rates of the state variables, but is defined through the
minimum dissipation of velocity fields that generate a prescribed admissible
tangent direction.

The state space is the transport manifold
\begin{equation}
\mathcal M
:=
\left\{
(\bfalpha,\mathbold R):
\divergence\bfalpha=\bfzero,
\qquad
\curl\mathbold R=\bfzero,
\qquad
\bfbeta_p=\mathcal C[\bfalpha]+\mathbold R
\right\}.
\label{eq:transport_manifold_alpha_R}
\end{equation}

We work formally on the bounded, smooth, simply connected domain $\Omega$ with connected boundary, using the decomposition and boundary conditions specified in Section~\ref{sec:kinematics}. We take $\bfalpha,\mathbold R\in L^2(\Omega)$, with $\mathbold R\in H^1(\Omega)$ when $W_R\neq0$, and impose the divergence and curl constraints in the distributional sense. The derivations assume sufficiently smooth fields, finite energy, and transport velocities of finite dissipation.
The boundary terms in the integrations by parts vanish under the conditions of Section~\ref{subsubsec:boundary_conditions}.
Although the compatibility constraints defining $\mathcal M$ are linear, the transport-generated directions generally do not span its full tangent space. Consequently, the evolution is restricted to states reachable from the initial state by transport.

At a state \((\bfalpha,\mathbold R)\in\mathcal M\), we denote by
\(T^{\mathrm{tr}}_{(\bfalpha,\mathbold R)}\mathcal M\) the
transport-admissible tangent space to \(\mathcal M\) at
\((\bfalpha,\mathbold R)\), where \(T\) denotes the tangent space and the
superscript \(\mathrm{tr}\) indicates restriction to transport-generated
directions. It is defined by
\begin{equation}
T^{\mathrm{tr}}_{(\bfalpha,\mathbold R)}\mathcal M
:=
\left\{
(s_{\bfalpha},s_R):
\begin{array}{l}
s_{\bfalpha}
=
\curl
\left(
\bfalpha\times\mathbold V_{\bfalpha}
\right),
\\[4pt]
\left(s_R\right)_{ij}
=
{\left(\mathcal P[\bfalpha\times\mathbold V_{\bfalpha}]\right)_{ij}}
-
\partial_j
\left(
R_{ik}V_R^k
\right)
\end{array}
\text{ for some }
(\mathbold V_{\bfalpha},\mathbold V_R)
\right\}.
\label{eq:transport_tangent_space_alpha_R}
\end{equation}
Here, \(\mathbold V_{\bfalpha}\) is the transport velocity associated with the
evolution of the dislocation density tensor governed by the dislocation transport law, while \(\mathbold V_R\) is the
transport velocity associated with the compatible slip \(\mathbold R\). Thus
\(s_{\bfalpha}\) and \(s_R\) are not arbitrary variations; they are precisely
the tangent directions generated by the admissible transport velocities.

Each admissible tangent direction
\((s_{\bfalpha},s_R)\) induces the plastic-distortion rate \(\bfeta\) given in
\eqref{eq:eta_alpha_R_lie}. Thus, \(\bfeta\) is not an independent tangent
direction on \(\mathcal M\), but is induced by the admissible transport
velocities associated with \((s_{\bfalpha},s_R)\).
We assume that the mobility tensors
\(\mathbold M_{\bfalpha}\) and \(\mathbold M_R\) are symmetric positive
definite. We define the corresponding transport metric by assigning a kinetic cost to the
transport velocities. For an admissible tangent direction
\((s_{\bfalpha},s_R)\), set
\begin{equation}
\begin{aligned}
g_{(\bfalpha,\mathbold R)}
\left(
(s_{\bfalpha},s_R),
(s_{\bfalpha},s_R)
\right)
:=
\inf
\int_\Omega
\left[
\mathbold V_{\bfalpha}\cdot
\mathbold M_{\bfalpha}^{-1}
\mathbold V_{\bfalpha}
+
\mathbold V_R\cdot
\mathbold M_R^{-1}
\mathbold V_R
\right]
\,\mathrm d\Omega .
\end{aligned}
\label{eq:transport_metric_alpha_R}
\end{equation}
The infimum is taken over all velocity fields
\((\mathbold V_{\bfalpha},\mathbold V_R)\) satisfying
\begin{equation}
s_{\bfalpha}
=
\curl
\left(
\bfalpha\times\mathbold V_{\bfalpha}
\right),
\qquad
\left(s_R\right)_{ij}
=
{\left(\mathcal P[\bfalpha\times\mathbold V_{\bfalpha}]\right)_{ij}}
-
\partial_j
\left(
R_{ik}V_R^k
\right) .
\label{eq:metric_velocity_constraints_alpha_R}
\end{equation}
Equivalently, the metric measures the least dissipation required to realize a
given admissible tangent direction. This is the direct analog of the
Wasserstein metric, where a density tangent direction is measured by the least
kinetic energy of a velocity field producing it through the continuity
equation. Here the admissible tangent directions are generated instead by
the transport law for \(\bfalpha\) and by the {evolution law
\eqref{eq:R_lie_transport}} for the compatible slip \(\mathbold R\).

We now make precise the sense in which the constrained Onsager evolution can
be interpreted as a gradient flow of the reduced free energy
\(\mathcal F^{\mathrm{red}}\) with respect to the transport metric
\eqref{eq:transport_metric_alpha_R}. The following result is formal and is
intended in the formal Riemannian sense described above: admissible tangent directions are
generated by transport velocities, and the metric assigns to each tangent
direction the minimum dissipation among the velocity fields that realize it.
Equivalently, velocity fields that generate the same tangent direction are
identified through their minimum-dissipation representative.

Throughout this subsection, we assume sufficient regularity of the state and
velocity fields, and the existence of the required minimum-dissipation
representatives.
The rigorous proof by Otto for mass transport in porous media \cite{Otto2000} uses special properties of the scalar Wasserstein/porous-medium problem, including approximation and compactness results that we have not established for our transport system governed by the dislocation transport law. 
Our transport map is more complicated and has a nontrivial kernel, so we would need results for the induced metric, weak solutions, and convergence.
Consequently, a complete metric-space theory, including the rigorous construction of the induced transport distance and a corresponding weak-solution framework, is beyond the scope of the present work.

The integrations by parts are understood under the boundary conditions of Section~\ref{subsubsec:boundary_conditions}.

\begin{proposition}[Formal gradient-flow identity]
\label{prop:transport_gradient_flow_identity}
Let \((\bfalpha(t),\mathbold R(t))\in\mathcal M\) be a smooth solution of the
transport-constrained evolution system \eqref{eq:closed_evolution_alpha_R_new}, where the
driving forces \(\mathbold f_{\bfalpha}\) and \(\mathbold f_R\) are defined by
\eqref{eq:alpha_force_general_new} and \eqref{eq:R_force_general_new}.
Assume that the mobility tensors
\(\mathbold M_{\bfalpha}\) and \(\mathbold M_R\) are symmetric positive
definite.

Then, for every admissible tangent direction
\begin{equation}
(s_{\bfalpha},s_R)
\in
T^{\mathrm{tr}}_{(\bfalpha,\mathbold R)}\mathcal M,
\label{eq:proposition_admissible_direction}
\end{equation}
the constrained evolution satisfies
\begin{equation}
g_{(\bfalpha,\mathbold R)}
\left(
(\dot{\bfalpha},\dot{\mathbold R}),
(s_{\bfalpha},s_R)
\right)
+
D\mathcal F^{\mathrm{red}}(\bfalpha,\mathbold R)
[
s_{\bfalpha},s_R
]
=
0 .
\label{eq:proposition_gradient_flow_identity}
\end{equation}
Consequently, for time-independent external loading,
\begin{equation}
\frac{\mathrm d}{\mathrm dt}
\mathcal F^{\mathrm{red}}[\bfalpha,\mathbold R]
=
-
\int_\Omega
\mathbold f_{\bfalpha}\cdot
\mathbold M_{\bfalpha}\mathbold f_{\bfalpha}
\,\mathrm d\Omega
-
\int_\Omega
\mathbold f_R\cdot
\mathbold M_R\mathbold f_R
\,\mathrm d\Omega
\le 0 .
\label{eq:proposition_energy_dissipation}
\end{equation}
Thus, in the formal Riemannian sense induced by the transport metric,
\begin{equation}
(\dot{\bfalpha},\dot{\mathbold R})
=
-
\operatorname{grad}_{g}
\mathcal F^{\mathrm{red}}(\bfalpha,\mathbold R).
\label{eq:proposition_gradient_flow_conclusion}
\end{equation}
\end{proposition}

\begin{proof}
Let
\begin{equation}
(s_{\bfalpha},s_R)
\in
T^{\mathrm{tr}}_{(\bfalpha,\mathbold R)}\mathcal M
\label{eq:proof_admissible_direction}
\end{equation}
be an arbitrary admissible tangent direction. By definition, there exist test
transport velocities
\((\mathbold W_{\bfalpha},\mathbold W_R)\) such that
\begin{equation}
s_{\bfalpha}
=
\curl
\left(
\bfalpha\times\mathbold W_{\bfalpha}
\right),
\qquad
\left(s_R\right)_{ij}
=
{\left(\mathcal P[\bfalpha\times\mathbold W_{\bfalpha}]\right)_{ij}}
-
\partial_j
\left(
R_{ik}W_R^k
\right) .
\label{eq:test_tangent_direction_alpha_R_proof}
\end{equation}
Here, \(s_R\) is the tangent direction for the compatible slip \(\mathbold R\).
Since \(\mathbold R\) itself enters the reconstruction
\(\bfbeta_p=\mathcal C[\bfalpha]+\mathbold R\), no additional kinematic tangent
direction \(\nabla s_R\) is introduced. Gradients of \(s_R\) enter only through
the variation of the regularizing energy \(W_R(\nabla\mathbold R)\), when such
an energy is included. When several velocity fields generate the same tangent
direction, we take \((\mathbold W_{\bfalpha},\mathbold W_R)\) to be the
minimum-dissipation representative appearing in the definition of the metric
\eqref{eq:transport_metric_alpha_R}. With this convention, the metric pairing
is independent of the representative.

Using the force representation of the reduced-energy variation, the
directional derivative of \(\mathcal F^{\mathrm{red}}\) along
\((s_{\bfalpha},s_R)\) is
\begin{equation}
D\mathcal F^{\mathrm{red}}(\bfalpha,\mathbold R)
[
s_{\bfalpha},s_R
]
=
-
\int_\Omega
\mathbold W_{\bfalpha}\cdot\mathbold f_{\bfalpha}
\,\mathrm d\Omega
-
\int_\Omega
\mathbold W_R\cdot\mathbold f_R
\,\mathrm d\Omega .
\label{eq:proof_energy_variation_test_tangent}
\end{equation}
The lattice energy contribution is included in this force representation
through the effective thermodynamic stress
\(\bfsigma^{\mathrm{eff}}\) defined in
\eqref{eq:effective_stress_with_lattice_energy}, which enters both
\(\mathbold f_{\bfalpha}\) and \(\mathbold f_R\).

On the other hand, the constrained Onsager principle gives the optimal
velocities in \eqref{eq:optimal_velocities_force_law_new}. Substitution of
these velocities into the transport laws gives the actual tangent direction
in \eqref{eq:closed_transport_evolution_alpha_R}.

Using the bilinear form induced by the transport metric and the
minimum-dissipation velocity representatives, we obtain
\begin{equation}
g_{(\bfalpha,\mathbold R)}
\left(
(\dot{\bfalpha},\dot{\mathbold R}),
(s_{\bfalpha},s_R)
\right)
+
D\mathcal F^{\mathrm{red}}(\bfalpha,\mathbold R)
\left[
s_{\bfalpha},s_R
\right]
=
\int_\Omega
\mathbold W_{\bfalpha}\cdot
\left(
\mathbold M_{\bfalpha}^{-1}
\mathbold V_{\bfalpha}^*
-
\mathbold f_{\bfalpha}
\right)
\,\mathrm d\Omega
+
\int_\Omega
\mathbold W_R\cdot
\left(
\mathbold M_R^{-1}
\mathbold V_R^*
-
\mathbold f_R
\right)
\,\mathrm d\Omega .
\label{eq:proof_gradient_flow_calculation}
\end{equation}
By \eqref{eq:optimal_velocities_force_law_new}, the right-hand side vanishes.
\begin{equation}
g_{(\bfalpha,\mathbold R)}
\left(
(\dot{\bfalpha},\dot{\mathbold R}),
(s_{\bfalpha},s_R)
\right)
+
D\mathcal F^{\mathrm{red}}(\bfalpha,\mathbold R)
[
s_{\bfalpha},s_R
]
=
0 .
\end{equation}
This proves \eqref{eq:proposition_gradient_flow_identity}.

It remains to derive the energy-dissipation identity. Along smooth solutions
and for time-independent prescribed loads, the chain rule gives
\begin{equation}
\frac{\mathrm d}{\mathrm dt}
\mathcal F^{\mathrm{red}}[\bfalpha,\mathbold R]
=
D\mathcal F^{\mathrm{red}}(\bfalpha,\mathbold R)
[
\dot{\bfalpha},\dot{\mathbold R}
] .
\label{eq:proof_energy_rate_chain_rule}
\end{equation}
Taking
\begin{equation}
(s_{\bfalpha},s_R)
=
(\dot{\bfalpha},\dot{\mathbold R})
\label{eq:proof_actual_direction}
\end{equation}
in \eqref{eq:proposition_gradient_flow_identity}, we obtain
\begin{equation}
\frac{\mathrm d}{\mathrm dt}
\mathcal F^{\mathrm{red}}[\bfalpha,\mathbold R]
=
-
g_{(\bfalpha,\mathbold R)}
\left(
(\dot{\bfalpha},\dot{\mathbold R}),
(\dot{\bfalpha},\dot{\mathbold R})
\right).
\label{eq:proof_energy_rate_metric_form}
\end{equation}
Using the optimal velocities
\eqref{eq:optimal_velocities_force_law_new} gives precisely the
energy-dissipation identity
\eqref{eq:proposition_energy_dissipation}.
\end{proof}
The proposition shows that the constrained Onsager system is a gradient flow with respect to the transport metric on the state variables \((\bfalpha,\mathbold R)\). 
The role of the metric is to measure the least
dissipation required to generate an admissible tangent direction through the
transport velocities. Thus, the velocity fields are the variables selected by
the Onsager principle, while the state rates
\((\dot{\bfalpha},\dot{\mathbold R})\) are induced by the transport laws. 

\section{Dislocation Driving Force and Reduction to the Classical Peach--Koehler Limit}
\label{sec:PK}

The dislocation driving force in \eqref{eq:alpha_force_general_new} differs from the classical Peach--Koehler force on a dislocation; the elastic terms correspond to the classical expression, but the defect energy and the lattice energy provide -- as expected -- additional contributions.
Importantly, these contributions are not appended independently to the transport law but instead follow from the variation of the energy along transport-admissible directions. 

The defect energy contribution to the dislocation driving force is:
\begin{equation}
    \mathbold f_{\bfalpha}^{\mathrm{def}}
    =
    \sum_i
    \left[
    -\curl\mathbold P_{\alpha,i}^{\mathrm d}
    \right]
    \times\bfalpha_i .
    \label{eq:defect_only_force}
\end{equation}
Related defect-energy, back-stress, line-tension, gradient, and curvature contributions arise in continuum dislocation theories \cite{GromaCsikorZaiser2003,Zaiser2015,Hochrainer2016,clayton2015defects}.


To consider the lattice energy contribution, we first neglect the defect and compatible slip regularizing energies, i.e., $W_\alpha = W_R = 0$.
The driving forces for $\bfalpha$ and $\bfR$ reduce to:
\begin{subequations}
\begin{align}
    \mathbold f_{\bfalpha}
    & =
    \sum_i
    \bfsigma_i^{\mathrm{eff}}
    \times
    \bfalpha_i .
    \label{eq:alpha_force_simple_new}
\\
    \left(
    \mathbold f_R
    \right)_k
    & =
    R_{ik}\partial_j\bfsigma_{ij}^{\mathrm{eff}}.
    \label{eq:R_force_simple_new}
\end{align}
\end{subequations}
Consequently, when the defect and compatible slip regularizing energies are neglected, the constrained Onsager principle yields a generalized Peach--Koehler-type transport force.
This generalized Peach--Koehler force has the same structure as the classical Peach--Koehler force, but with the effective thermodynamic stress $\bfsigma^{\mathrm{eff}}$ replacing the standard elastic stress. 
The compatible slip $\mathbold R$ is driven by its coupling to the effective thermodynamic stress through the
reconstructed plastic distortion.

\begin{remark}[Recovery of the Classical Peach--Koehler Force for a Single Dislocation Line] 
We show that the generalized dislocation driving force reduces to the classical Peach--Koehler force for a single dislocation line. 
Consider a dislocation with Burgers vector $\bfb$ and line direction $\bfxi$.
Its dislocation density tensor is
\begin{equation}
\bfalpha
=
\bfb\otimes\bfxi\,\delta_L,
\label{eq:single_dislocation_alpha_new}
\end{equation}
where $\delta_L$ denotes concentration on the dislocation line $L$. Thus,
row-wise,
\begin{equation}
\bfalpha_i
=
b_i\bfxi\,\delta_L .
\label{eq:single_dislocation_alpha_row_new}
\end{equation}
Substitution into \eqref{eq:alpha_force_simple_new} gives the force per unit
length
\begin{equation}
\mathbold f_{\mathrm{line}}^{\mathrm{eff}}
=
\sum_i
b_i
\left(
\bfsigma_i^{\mathrm{eff}}
\times
\bfxi
\right)
=
\left[
\left(
\bfsigma^{\mathrm{eff}}
\right)^T
\bfb
\right]
\times
\bfxi .
\label{eq:line_force_from_alpha_new}
\end{equation}
This is a generalized Peach--Koehler-type force in which the lattice energy contributes to the elastic part.
We highlight that, in general, $\bfsigma^{\mathrm{eff}}$ may not be symmetric because $\parderiv{\Psi}{\bfbeta_p}$ may not be symmetric; therefore, the transpose in \eqref{eq:line_force_from_alpha_new} cannot generally be omitted.

If we further neglect the lattice-energy contribution, i.e., $\parderiv{\Psi}{\bfbeta_p}=\bfzero$, then $\bfsigma^{\mathrm{eff}}=\bfsigma^*$, and we recover the classical expression for the Peach--Koehler force on a dislocation line:
\begin{equation}
\mathbold f_{\mathrm{line}}
=
\left(
\bfsigma^*
\bfb
\right)
\times
\bfxi ,
\label{eq:classical_PK_recovered_new}
\end{equation}
where we can now use the symmetry of the stress tensor.
See, for example, Ref.~\cite{PeachKoehler2020} for a modern discussion of driving forces on dislocations.
\end{remark}

\section{Comparison with Cahn--Hilliard-type Phase Field Evolution}
\label{sec:comparison_cahn_hilliard}

Phase-field and mesoscale formulations provide another important class of continuum descriptions of dislocation microstructure. 
Approaches developed by Koslowski, LeSar, and Thomson~\cite{LeSar2004}, LeSar~\cite{LeSar2014}, and Beyerlein and Hunter~\cite{BeyerleinHunter2016} illustrate how dislocation microstructure, plastic response, and defect interactions can be represented across scales. 
Such models motivate comparison with the present formulation, while also emphasizing that the choice of order parameter, conservation law, and admissible kinetics is central to the resulting evolution.

Within this broader class, Cahn--Hilliard dynamics provides a particularly useful comparison because it is a gradient flow with a conserved quantity \cite{CahnHilliard1958,Cahn1961}, and can also be expressed through a flux-based Rayleigh principle. 
The essential distinction lies in the underlying conservation law and admissible current: the relevant kinematics here are generated by the dislocation transport law rather than by the standard continuity equation as in the Cahn--Hilliard model.

For a conserved scalar order parameter $q$, the evolution may be written as
\begin{equation}
    \dot q
    +
    \divergence\mathbold J
    =
    0,
    \qquad
    \mathbold J
    =
    -
    M\nabla\mu,
    \qquad
    \mu
    =
    \frac{\delta\mathcal F}{\delta q}
    \implies
        \dot q
    =
    \divergence\left(
        M\nabla\mu
    \right),
\label{eq:cahn_hilliard_flux_form}
\end{equation}
where $\mathbold J$ is the flux, $M$ is the mobility, and $\mu$ is the
chemical potential which is the variational derivative of the free energy $\mathcal F$ with respect to $q$. 
This equation expresses the conservation of $q$ through an ordinary
diffusive flux driven by gradients of the chemical potential. For constant
mobility, it is the $H^{-1}$ gradient flow of the free energy; more general
mobilities use the corresponding weighted $H^{-1}$-type structures \cite{Carter1997}.

The same evolution can also be derived using an Onsager functional. Using
$\dot q=-\divergence\mathbold J$, the free-energy rate is
\begin{equation}
    \frac{\mathrm d}{\mathrm dt}\mathcal F[q]
    =
    \int_\Omega
    \mu\,\dot q
    \,\mathrm d\Omega
    =
    \int_\Omega
    \nabla\mu\cdot\mathbold J
    \,\mathrm d\Omega,
\label{eq:cahn_hilliard_energy_rate_flux}
\end{equation}
where {the boundary term vanishes by the no-flux condition of Section~\ref{subsubsec:boundary_conditions}}. Minimizing the Rayleighian
\begin{equation}
    \mathcal R_{\mathrm{CH}}[\mathbold J]
    :=
    \frac{1}{2}
    \int_\Omega
    \mathbold J\cdot M^{-1}\mathbold J
    \,\mathrm d\Omega
    +
    \int_\Omega
    \nabla\mu\cdot\mathbold J
    \,\mathrm d\Omega
\label{eq:cahn_hilliard_rayleighian}
\end{equation}
with respect to the flux gives
$\mathbold J=-M\nabla\mu$ and hence recovers
\eqref{eq:cahn_hilliard_flux_form}. Thus, the essential
Cahn--Hilliard procedure consists of combining the appropriate conservation
law with a free energy and a dissipation potential for the corresponding
flux.

The corresponding conservation law for the dislocation
density tensor is \eqref{eq:alpha_tangent_direction}.
Recalling the dislocation flux introduced in
Section~\ref{subsec:lie_derivative_alpha}, we have:
\begin{equation}
    \mathbold J_\alpha
    :=
    \bfalpha\times\mathbold V_\alpha
\label{eq:mura_current_definition}
\end{equation}
The requirement that the flux has the geometric structure above is a key distinction compared to the Cahn--Hilliard flux in \eqref{eq:cahn_hilliard_flux_form}.
Consequently, the resulting negative-order metric for dislocation evolution is not a fixed \(H^{-1}\) norm as in the case of Cahn--Hilliard evolution.
Because the transport operator associated with the dislocation transport law \(\curl(\bfalpha\times\mathbold V_\alpha)\) depends explicitly on the current dislocation density, the corresponding metric is state-dependent and is precisely the transport metric introduced in Section~\ref{subsec:transport_manifold_metric}.

The Onsager functional approach can similarly be adapted by replacing the standard flux with the dislocation flux that is compatible with the dislocation transport law, and minimizing over the dislocation velocity (Section~\ref{sec:constrained_gradient_flow_dislocation_evolution}).

Recent phase-field interpretations of continuum dislocation dynamics have
likewise emphasized that dislocation-density variables obey nonstandard
conservation laws and that their evolution equations generally go beyond the
classical distinction between Allen--Cahn and Cahn--Hilliard dynamics
\cite{Yufan2025}. For curved dislocations, the evolution of line density may
contain transport, orientation, and curvature contributions, and the total
line length need not be conserved.

{
\section{Application to Dislocation Annihilation}
\label{sec:examples}

We now specialize the general formulation to the evolution of two dislocations with equal and opposite Burgers charge, and study their annihilation dynamics.
Our aim is to understand how the elastic, defect, nonconvex lattice, and applied stress contributions modify the dislocation interactions and dynamics.

We restrict the continuum fields to a single
glide layer, on which they are described by a
signed density $q(x_1,t)$.
The transport law and the Onsager principle then give
an evolution equation for $q$, which we solve
numerically from Gaussian initial cores.
We focus on narrow cores, for which the lattice energy
is large compared with the elastic interaction at the
core scale, as near closely spaced dislocation
cores~\cite{Hu2020}.
The cores then keep the Peierls--Nabarro width and
glide toward annihilation, and a uniform resolved shear
stress either assists annihilation or, above a
threshold, separates the pair.

The calculations are illustrative rather than quantitative
predictions for a specific material: the parameters have not been
calibrated to a particular crystal, although first-principles calculations
provide a route to such calibration~\cite{Behnoud2025}.


\subsection{Geometry and Modeling Assumptions}
\label{subsec:collective_coordinate_reduction}

\input{media/jmps_section7_schematic}

We consider two straight, parallel edge dislocations
with Burgers vectors $\pm q_0\mathbold e_1$, $q_0>0$,
line direction $\mathbold e_3$, and common glide
plane $x_2=0$.
The body is large compared with the initial separation
$d_0$ and the core width,
so that boundary and image effects are neglected,
integrals along $x_1$ are taken over $\mathbb R$,
and all energies are per unit dislocation-line length.
Figure~\ref{fig:opposite_pair_geometry}
illustrates the geometry and the evolution of the
density.
We make the following assumptions:
\begin{description}
    \item[(i) Glide-layer kinematics]
    
    Slip is confined to a single glide layer
    $|x_2|<h/2$ of interplanar thickness $h$.
    The dislocation density and the compatible slip are
    \begin{equation}
    \begin{aligned}
    \bfalpha(\bfx,t)
    &=
    \frac{q(x_1,t)}{h}\,
    \chi_h(x_2)\,
    \mathbold e_1\otimes\mathbold e_3,
    \\
    \mathbold R(\bfx,t)
    &=
    \frac{\gamma(x_1,t)}{h}\,
    \chi_h(x_2)\,
    \mathbold e_1\otimes\mathbold e_2
    -\mathcal C[\bfalpha],
    \end{aligned}
    \label{eq:opposite_pair_alpha_reduction}
    \end{equation}
    where $\chi_h$ is the indicator function of the
    layer, $q$ is the signed Burgers content per unit
    length of the glide plane, and
    $\gamma(x_1,t):=\int_{-\infty}^{x_1}q(x_1',t)\,\mathrm dx_1'$
    is the slip, so that $q=\partial_{x_1}\gamma$.
    
    Then $\divergence\bfalpha=\bfzero$, and
    $\curl\mathbold R=\bfzero$ because
    $\curl\bigl((\gamma/h)\,\chi_h\,\mathbold e_1\otimes\mathbold e_2\bigr)=\bfalpha$
    and, by \eqref{eq:div_curl_problem_C},
    $\curl\mathcal C[\bfalpha]=\bfalpha$.

    The plastic distortion
    $\bfbeta_p=\mathcal C[\bfalpha]+\mathbold R$
    from \eqref{eq:beta_p_reconstruction_notation}
    is then the slip on the layer:
    \begin{equation}
    \bfbeta_p(\bfx,t)
    =
    \frac{\gamma(x_1,t)}{h}\,
    \chi_h(x_2)\,
    \mathbold e_1\otimes\mathbold e_2.
    \label{eq:opposite_pair_beta_p_scalar_embedding}
    \end{equation}
    We set $W_R=0$ without imposing
    $\mathbold R=\bfzero$.
    The energy then depends on $\mathbold R$ only
    through $\bfbeta_p$, which is determined by $\gamma$,
    so the evolution of $\mathbold R$ need not be computed
    separately (described in (ii) below).

    \item[(ii) Transport] 

    For glide, $\mathbold V_{\bfalpha}=V(x_1,t)\,\mathbold e_1$,
    where $V$ depends on the whole profile $q(\cdot,t)$ and
    is determined by the Onsager principle
    (Section~\ref{subsubsec:reduced_onsager_evolution}),
    and the transport law \eqref{eq:alpha_tangent_direction}
    reduces to
    \begin{equation}
    \dot q=\partial_{x_1}(qV),
    \label{eq:glide_layer_transport}
    \end{equation}
    or, integrating with $\gamma(-\infty,t)=0$,
    $\dot\gamma=qV$.
    We set $\mathbold V_R=\bfzero$, the limit of vanishing
    compatible-slip mobility $\mathbold M_R$, so that
    $\mathbold R$ changes only through the compatible slip
    generated by dislocation motion.
    Then \eqref{eq:R_lie_transport} and
    \eqref{eq:eta_alpha_R_lie} give
    $\dot\bfbeta_p=\bfalpha\times\mathbold V_{\bfalpha}
    =(qV/h)\,\chi_h\,\mathbold e_1\otimes\mathbold e_2$
    and $\dot{\mathbold R}=\mathcal P[\dot\bfbeta_p]$\; the plastic-distortion rate is then that of Kosevich and Mura~\cite{Kosevich1962,mura1963continuous}.
    Hence $\bfbeta_p$ keeps the form
    \eqref{eq:opposite_pair_beta_p_scalar_embedding}
    for all $t$, and the state is described by $\gamma$ alone.

    \item[(iii) Drag mobility]
    We take the glide mobility to be
    $m/|\bfalpha|$, with $m>0$ constant.
    The dissipation density
    $|\bfalpha|\,|\mathbold V_{\bfalpha}|^2/(2m)$
    then weights the velocity by the local Burgers
    content, as in the Wasserstein metric of
    Section~\ref{sec:constrained_gradient_flow_dislocation_evolution},
    and gives the classical linear drag law for a
    dislocation line~\cite{HirthLothe1982}.
    As shown in
    Section~\ref{subsubsec:reduced_onsager_evolution},
    the velocity then reverses direction where $q$ changes
    sign: positive and negative Burgers content move in
    opposite directions under the same stress.

    \item[(iv) Initial state]
    At $t=0$, the density consists of two Gaussian cores
    of width $\varepsilon\gg h$, centered at
    $x_1=\mp d_0/2$:
    \begin{equation}
    q(x_1,0)
    =
    q_0\left[
    G_\varepsilon\left(x_1+\frac{d_0}{2}\right)
    -
    G_\varepsilon\left(x_1-\frac{d_0}{2}\right)
    \right],
    \quad\text{where}\quad
    G_\varepsilon(s)
    :=
    \frac{1}{\sqrt{2\pi}\varepsilon}
    \exp\left(-\frac{s^2}{2\varepsilon^2}\right).
    \label{eq:opposite_pair_profile_ansatz}
    \end{equation}
    Each core carries Burgers content of magnitude
    $q_0$, with opposite signs.
    For $t>0$, $q$ evolves by
    \eqref{eq:glide_layer_transport}, and the shape and
    width of the cores change with time.
    Since the initial state and the loading are symmetric
    about $x_1=0$, $q(-x_1,t)=-q(x_1,t)$ for all $t$, and
    $q$ changes sign at $x_1=0$.
    Since \eqref{eq:glide_layer_transport} transports $q$,
    it preserves the sign of $q$ along particle paths, so
    $q\geq0$ for $x_1<0$ and $q\leq0$ for $x_1>0$.
    
    \item[(v) Annihilation]
    By (iii), the velocity has opposite directions on the
    two sides of $x_1=0$.
    When the net stress drives the cores together,
    positive and negative Burgers content are carried
    toward $x_1=0$ and cancel there, and the slip
    $\gamma(0,t)$, which equals the Burgers content on
    either side of $x_1=0$, decreases to zero.
    Annihilation thus follows from the transport law.
    At $x_1=0$, the velocity need not be Lipschitz continuous\footnote{
    {Since a continuous velocity conserves the Burgers vector of every convected circuit, annihilation requires a velocity that is not Lipschitz continuous where opposite-sign Burgers content meets. Alternatively, the opposite-sign Burgers dislocations can be considered as distinct populations of dislocations, each with its own velocity \cite{Monavari2018}.}
    }, and
    \eqref{eq:glide_layer_transport} is understood in the
    flux form $\dot\gamma=qV$.
    
\end{description}

\subsection{Energies and Evolution Equation}
\label{subsec:reduced_pair_energy_evolution}

The energy is the free energy
\eqref{eq:reduced_free_energy_alpha_R} evaluated on
\eqref{eq:opposite_pair_alpha_reduction}--\eqref{eq:opposite_pair_beta_p_scalar_embedding},
up to constants; it is a functional of the slip
$\gamma$.
For each contribution, we give its variational
derivative with respect to $\gamma$, which determines
the driving force.

\subsubsection{Elastic Energy}

We take isotropic elasticity with shear modulus $\mu$
and Poisson's ratio $\nu$.
After minimizing out the displacement, the elastic
energy of \eqref{eq:opposite_pair_alpha_reduction}
is~\cite{Peierls1940,Nabarro1947,HirthLothe1982}:
\begin{equation}
\mathcal E_{\rm el}
:=
-\frac{K}{2q_0^2}
\int_{\mathbb R}\int_{\mathbb R}
q(x_1)\,q(x_1')\,
\log|x_1-x_1'|\,
\mathrm dx_1\,\mathrm dx_1',
\quad\text{where}\quad
K:=\frac{\mu q_0^2}{2\pi(1-\nu)}.
\label{eq:pn_elastic_energy}
\end{equation}
Using $\delta q=\partial_{x_1}\delta\gamma$ and
integrating by parts, its variational derivative is
\begin{equation}
\frac{\delta\mathcal E_{\rm el}}{\delta\gamma}(x_1)
=
\frac{K}{q_0^2}\,
\mathrm{p.v.}\!\int_{\mathbb R}
\frac{q(x_1')}{x_1-x_1'}\,\mathrm dx_1',
\label{eq:elastic_variational_derivative}
\end{equation}
where $\mathrm{p.v.}$ denotes the Cauchy principal value.

For two point dislocations with Burgers content
$\pm q_0$ at separation $d$,
\eqref{eq:pn_elastic_energy} and
\eqref{eq:elastic_variational_derivative} recover the
classical energy $K\log d$ and attraction $K/d$ of two
coplanar opposite-sign edge
dislocations~\cite{HirthLothe1982}.
For smooth densities, such as the initial state
\eqref{eq:opposite_pair_profile_ansatz},
\eqref{eq:elastic_variational_derivative} is bounded:
the cores regularize the interaction.

\subsubsection{Core Energy}

We take the quadratic defect energy
$W_\alpha(\bfalpha)=\kappa|\bfalpha|^2/2$,
$\kappa\geq0$.
Since $|\bfalpha|^2=q^2\chi_h/h^2$,
\begin{equation}
\mathcal E_\alpha
:=
\frac{\kappa_h}{2}
\int_{\mathbb R}
q^2\,\mathrm dx_1,
\quad\text{where}\quad
\kappa_h:=\frac{\kappa}{h},
\quad\text{and}\quad
\frac{\delta\mathcal E_\alpha}{\delta\gamma}
=
-\kappa_h\,\partial_{x_1}q.
\label{eq:opposite_pair_reduced_defect_energy_definition}
\end{equation}
This contribution penalizes concentrated density and
therefore favors spreading of the cores.

\subsubsection{Lattice Energy}

On the glide layer,
\eqref{eq:opposite_pair_beta_p_scalar_embedding}
gives exactly
\begin{equation}
\int_\Omega\Psi(\bfbeta_p)\,\mathrm d\Omega
=
\int_{\mathbb R}\psi(\gamma)\,\mathrm dx_1,
\quad\text{where}\quad
\psi(\gamma)
:=
h\,\Psi\left(\frac{\gamma}{h}\,
\mathbold e_1\otimes\mathbold e_2\right),
\label{eq:glide_layer_lattice_energy}
\end{equation}
with $\Psi(\bfzero)=0$.
A slip $\gamma=b_{\rm slip}$ across the layer is the
lattice-equivalent increment
$\mathbold S=(b_{\rm slip}/h)\,\mathbold e_1\otimes\mathbold e_2$,
so the invariance of $\Psi$ assumed in
Section~\ref{sec:energetics} gives
$\psi(\gamma+b_{\rm slip})=\psi(\gamma)$.
We take the simplest such form:
\begin{equation}
\psi(\gamma)
=
A_\psi
\sin^2\left(
\frac{\pi\gamma}{b_{\rm slip}}
\right),
\qquad A_\psi>0.
\label{eq:opposite_pair_nonconvex_lattice_density}
\end{equation}
The lattice-equivalent minima occur at
$\gamma=n b_{\rm slip}$, $n\in\mathbb Z$.
We take $b_{\rm slip}=q_0$, so that the slip
plateau between sufficiently separated cores
approaches a lattice-energy minimum.

The lattice energy and its variational derivative are
\begin{equation}
\mathcal E_\psi
:=
\int_{\mathbb R}
\psi(\gamma)\,\mathrm dx_1,
\qquad
\frac{\delta\mathcal E_\psi}{\delta\gamma}
=
\psi'(\gamma)
=
\frac{\pi A_\psi}{b_{\rm slip}}
\sin\left(
\frac{2\pi\gamma}{b_{\rm slip}}
\right).
\label{eq:opposite_pair_reduced_lattice_energy}
\end{equation}
The lattice energy favors narrow cores, since it
penalizes slip between its minima, while the elastic
energy favors spreading.
For an isolated dislocation without defect energy and
applied stress, the balance
$\delta(\mathcal E_{\rm el}+\mathcal E_\psi)/\delta\gamma=0$
gives the Peierls--Nabarro
core~\cite{Peierls1940,Nabarro1947}
\begin{equation}
\gamma(x_1)
=
q_0\left[
\frac12
+
\frac1\pi
\arctan\left(\frac{x_1}{\zeta}\right)
\right],
\quad\text{where}\quad
\zeta
:=
\frac{K}{2\pi A_\psi},
\label{eq:pn_core}
\end{equation}
so the core half-width $\zeta$ is set by the ratio of
the elastic and lattice energies.

\subsubsection{Applied Stress and Total Energy}

A uniform resolved shear stress $\tau$, applied
through the boundary tractions, contributes the
load term
$-\int_\Omega\tau\,(\bfbeta_p)_{12}\,\mathrm d\Omega$~\cite{mura2013micromechanics}:
\begin{equation}
\mathcal E_\tau
:=
-\tau
\int_{\mathbb R}\gamma\,\mathrm dx_1,
\qquad
\frac{\delta\mathcal E_\tau}{\delta\gamma}
=
-\tau.
\label{eq:opposite_pair_applied_stress_energy}
\end{equation}

\subsubsection{Onsager Evolution}
\label{subsubsec:reduced_onsager_evolution}

The total energy is
\begin{equation}
\mathcal E_{\rm tot}
:=
\mathcal E_{\rm el}
+
\mathcal E_\alpha
+
\mathcal E_\psi
+
\mathcal E_\tau,
\label{eq:opposite_pair_total_reduced_energy}
\end{equation}
and, by \eqref{eq:elastic_variational_derivative},
\eqref{eq:opposite_pair_reduced_defect_energy_definition},
\eqref{eq:opposite_pair_reduced_lattice_energy}, and
\eqref{eq:opposite_pair_applied_stress_energy},
\begin{equation}
\frac{\delta\mathcal E_{\rm tot}}{\delta\gamma}(x_1)
=
\frac{K}{q_0^2}\,
\mathrm{p.v.}\!\int_{\mathbb R}
\frac{q(x_1')}{x_1-x_1'}\,\mathrm dx_1'
-
\kappa_h\,\partial_{x_1}q
+
\psi'(\gamma)
-
\tau.
\label{eq:opposite_pair_total_reduced_force}
\end{equation}

With $\dot\gamma=qV$ from \eqref{eq:glide_layer_transport} and assumption~(ii), the energy
rate is
\begin{equation}
\frac{\mathrm d}{\mathrm dt}\mathcal E_{\rm tot}
=
\int_{\mathbb R}
\frac{\delta\mathcal E_{\rm tot}}{\delta\gamma}\,
qV\,\mathrm dx_1
=
-\int_{\mathbb R}fV\,\mathrm dx_1,
\quad\text{where}\quad
f
:=
q\,\tau_{\rm tot},
\quad\text{and}\quad
\tau_{\rm tot}
:=
-\frac{\delta\mathcal E_{\rm tot}}{\delta\gamma}.
\label{eq:pair_pointwise_force_law}
\end{equation}
Here $\tau_{\rm tot}$ is the total resolved shear
stress on the glide plane: by
\eqref{eq:opposite_pair_total_reduced_force}, the
applied stress $\tau$ plus the stresses from the
elastic interaction, the core energy, and the lattice
energy; and $f$ is the glide force per unit length
of the glide plane, which reduces to the Peach--Koehler
force~\cite{HirthLothe1982} when only the elastic and
applied stresses are retained.
Comparing with \eqref{eq:reduced_energy_rate_force_form_new},
$f$ is the integral of $(\mathbold f_{\bfalpha})_1$
across the glide layer.
With the drag mobility of assumption~(iii), the
dissipation density
$|\bfalpha|\,|\mathbold V_{\bfalpha}|^2/(2m)$
integrates across the layer to $qV^2/(2m)$ for
$x_1<0$, where $q\geq0$ by assumption~(iv).
By symmetry, the Onsager functional
\eqref{eq:onsager_functional_alpha_R} is twice
\begin{equation}
\mathcal J
=
\int_{-\infty}^{0}
\left(\frac{qV^2}{2m}-fV\right)\mathrm dx_1 .
\label{eq:reduced_pair_onsager}
\end{equation}
Its stationarity gives $qV/m=f=q\,\tau_{\rm tot}$, so
\begin{equation}
V
=
m\,\tau_{\rm tot},
\qquad x_1<0,
\label{eq:pair_velocities}
\end{equation}
and, by symmetry, $V=-m\,\tau_{\rm tot}$ for $x_1>0$.
Positive and negative Burgers content therefore move
with the same speed in opposite directions, as in
continuum dislocation dynamics with signed
densities~\cite{Monavari2018}.
Substituting \eqref{eq:pair_velocities} into
$\dot\gamma=qV$ gives the evolution equation
\begin{equation}
\dot\gamma
=
-m\,\partial_{x_1}\gamma\,
\frac{\delta\mathcal E_{\rm tot}}{\delta\gamma},
\qquad x_1<0,
\label{eq:general_reduced_pair_dynamics}
\end{equation}
with $\delta\mathcal E_{\rm tot}/\delta\gamma$ given by
\eqref{eq:opposite_pair_total_reduced_force} and
$\gamma(-x_1,t)=\gamma(x_1,t)$.
The core shapes evolve freely:
\eqref{eq:general_reduced_pair_dynamics} is the
continuum evolution
\eqref{eq:closed_evolution_alpha_R_new} on the glide
layer under assumptions (i)--(iii).
It is a nonlocal Hamilton--Jacobi equation for
$\gamma$; related nonlocal Hamilton--Jacobi systems
arise in dislocation-density
dynamics~\cite{ElHajjForcadel2008}. Without the lattice and core energies, \eqref{eq:general_reduced_pair_dynamics} has the form of the continuum limit of discrete dislocations with annihilation~\cite{vanMeursPeletierPozar2022}, and the dynamic Peierls--Nabarro model, in which $\gamma$ instead follows an $L^2$ gradient flow, converges to discrete dislocation dynamics with annihilation~\cite{vanMeursPatrizi}. Here, Peierls--Nabarro cores form and glide under the transport-constrained evolution.

At fixed applied stress, the evolution satisfies
\begin{equation}
\frac{\mathrm d}{\mathrm d t}\mathcal E_{\rm tot}
=
-2m\int_{-\infty}^{0}q
\left(\frac{\delta\mathcal E_{\rm tot}}{\delta\gamma}\right)^2
\mathrm dx_1
\leq0,
\label{eq:reduced_energy_decay}
\end{equation}
which is the glide-layer form of \eqref{eq:proposition_energy_dissipation}.
To compare the energetic contributions, we
introduce the dimensionless quantities
\begin{equation}
D_0:=\frac{d_0}{\varepsilon},
\qquad
\Lambda:=\frac{K}{A_\psi\varepsilon},
\qquad
\Gamma_\alpha
:=
\frac{\kappa_h q_0^2}{A_\psi\varepsilon^2},
\qquad
T:=\frac{\tau q_0}{A_\psi}.
\label{eq:pair_dimensionless_parameters}
\end{equation}
Here $D_0$ is the normalized initial separation,
$\Lambda$ and $\Gamma_\alpha$ measure
the elastic and defect contributions relative
to the lattice-energy scale, and $T$ is the
dimensionless applied stress.
By \eqref{eq:pn_core}, $\zeta/\varepsilon=\Lambda/(2\pi)$,
so $\Lambda$ also compares the Peierls--Nabarro core
width with the initial core width.

We solve \eqref{eq:general_reduced_pair_dynamics},
extended to $x_1>0$ by symmetry, numerically on a
uniform grid over an interval $[-L,L]$ with $L\gg d_0$.
The derivative $\partial_{x_1}\gamma$ is discretized with
an upwind scheme for Hamilton--Jacobi
equations~\cite{OsherSethian1988}. The elastic term is
the exact derivative of the energy
\eqref{eq:pn_elastic_energy}, discretized with the
logarithmic kernel averaged over grid cells, so the
principal value in \eqref{eq:elastic_variational_derivative}
is not evaluated directly; the convolution over the grid is evaluated exactly with the fast Fourier transform, after zero-padding to at least twice the number of grid points. The explicit time step is small enough for
stability, and the computed energy decreases
monotonically in all runs.


\subsection{Annihilation and Unbinding of a Dislocation Pair}
\label{subsec:applied_stress_trapping}

We consider cores for which the lattice energy is
large compared with the elastic interaction, so that
the Peierls--Nabarro core is much narrower than the
initial cores.
To isolate the effects of lattice energetics and
applied loading, we take a small core energy and use
\begin{equation}
\begin{gathered}
q_0=b_{\rm slip}=1,
\qquad
m=1,
\qquad
K=0.00625,
\qquad
\kappa_h=2\times10^{-6},
\\
\varepsilon=0.025,
\qquad
A_\psi=1,
\qquad
d_0=0.20,
\end{gathered}
\label{eq:opposite_pair_numerical_parameters}
\end{equation}
so that $\Lambda=0.25$, $\Gamma_\alpha=0.0032$, and $D_0=8$;
the applied stress $T$ is varied.
The initial separation of eight Gaussian core widths
keeps the initial core overlap negligible; the
calibration of such lengths depends on the underlying
continuum and atomistic
descriptions~\cite{Cai2006,Davoudi2020,Hu2020,Benzerga2005}.
Then $\zeta/\varepsilon=\Lambda/(2\pi)\approx0.040$, so
$\zeta\approx0.001$.
We solve on $[-1,1]$ with grid spacing
$1.25\times10^{-4}\approx\zeta/8$. The scheme converges at first order in the grid spacing: halving it reduces the annihilation time at $T=0$ by about $2\%$, and the time extrapolated to zero spacing is $1.59$, about $3\%$ less than the value $1.65$ on this grid.

Figure~\ref{fig:applied_stress_trapping} summarizes the
results at $T=0$.
The cores first sharpen, within $t\approx0.05$, from the
initial Gaussian profiles to nearly the Peierls--Nabarro
core \eqref{eq:pn_core}, about $11\%$ wider because of the
core energy
(Figs.~\ref{fig:applied_stress_trapping-a}
and~\ref{fig:applied_stress_trapping-b}).
They then glide toward each other with this shape
until they meet at $x_1=0$, where their Burgers content
cancels; $\gamma(0,t)$ reaches $q_0/2$ at
$t\approx1.65$. Two point dislocations moving with $V=m\,\tau_{\rm tot}$ under the attraction $K/d$ meet at $t=q_0d_0^2/(4mK)=1.6$, in agreement with the extrapolated time $1.59$. This velocity does not depend on the core width; in the dynamic Peierls--Nabarro model, the mobility of a core is instead inversely proportional to the integral of $q^2$, and hence proportional to its width~\cite{GonzalezMonneau2012}.
The energy of an isolated core does not depend on its
position, so the continuum lattice energy sets the core
width but does not resist core motion, and the elastic
attraction is unopposed.
The energies (Fig.~\ref{fig:applied_stress_trapping-en})
show three stages.
While the cores form, $\mathcal E_{\rm tot}$ drops
rapidly, from $0.088$ to $0.036$ by $t=0.05$: the
lattice energy decreases from $0.073$ to $0.008$, while
the elastic and core energies increase.
While the cores glide, $\mathcal E_{\rm tot}$ decreases
slowly, to $0.024$ at $t=1.6$; only the elastic energy
changes, since the lattice and core energies of a
gliding core are constant.
When the cores annihilate, all energies drop rapidly to
nearly zero.

An applied stress changes this picture
(Fig.~\ref{fig:regime_II_stress}), where
$s(t)$ denotes the distance between the centroids of
the positive and negative parts of $q$.
Negative $T$ drives the cores together and speeds up
annihilation: for $T=-0.05$, $\gamma(0,t)$ reaches
$q_0/2$ at $t\approx0.83$.
Positive $T$ pulls the cores apart.
For $T=0.02$, annihilation is delayed to
$t\approx3.06$; for $T=0.04$, the cores separate, and
$s$ grows to $1.6\,d_0$ by $t=4$.
Bisection on $T$ gives annihilation for
$T\leq0.03125$ and unbinding for $T\geq0.03163$,
which a finer grid confirms, so the threshold stress is
\begin{equation}
T_{\rm c}\approx0.0314
\approx
\frac{\Lambda}{D_0}
=
0.03125.
\label{eq:regime_II_threshold}
\end{equation}
The right-hand side is the force balance
$\tau q_0=K/d_0$ of two point dislocations at the
initial separation.
Since $\zeta\ll d_0$, the narrow cores interact as
point dislocations, and the threshold is set by the
elastic attraction at the initial separation.

\begin{figure*}[htbp]
    \centering
    \subfloat[]{%
        \label{fig:applied_stress_trapping-a}%
        \includegraphics[width=0.47\textwidth]{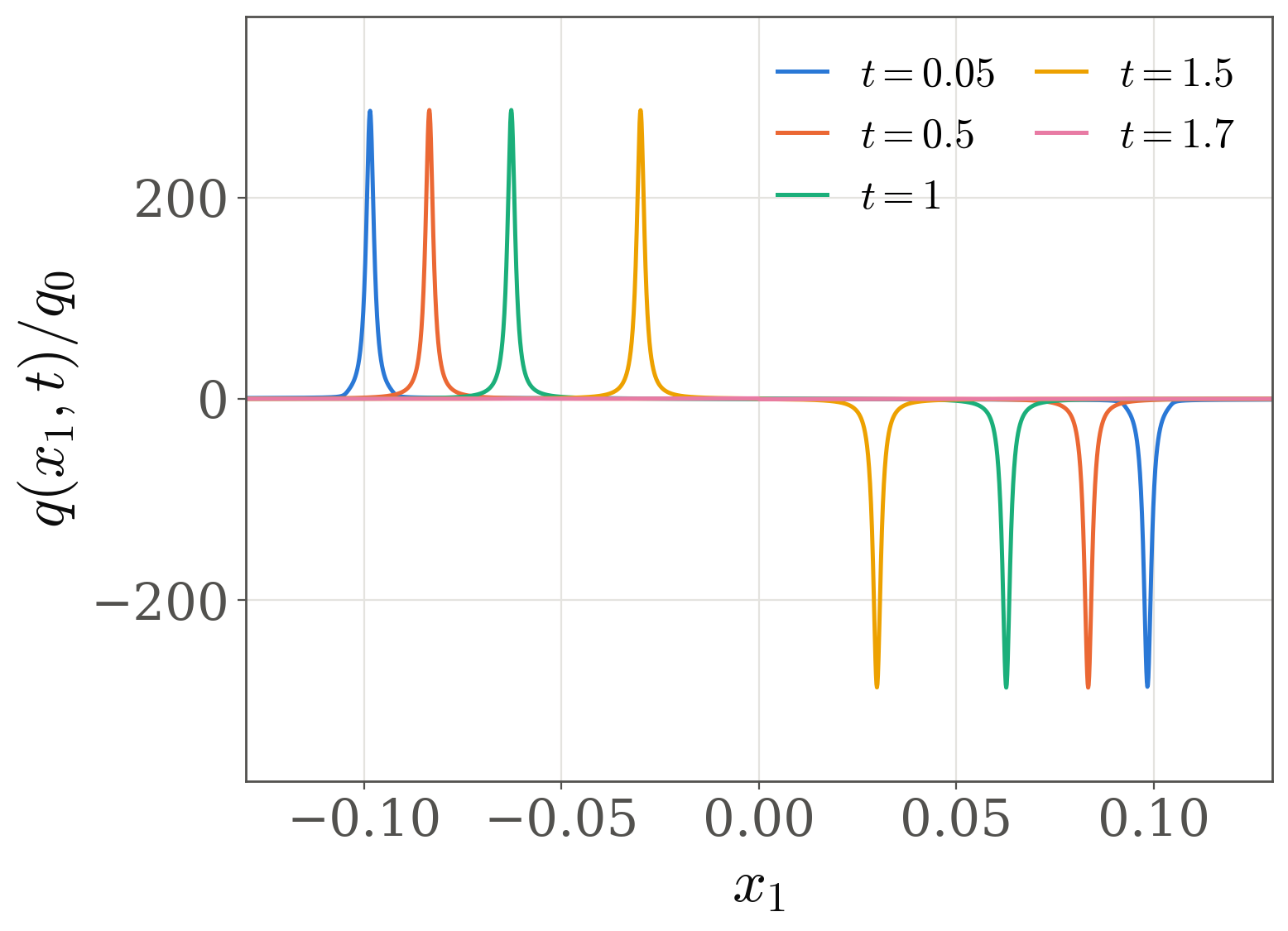}%
    }
    \hfill
    \subfloat[]{%
        \label{fig:applied_stress_trapping-en}%
        \includegraphics[width=0.47\textwidth]{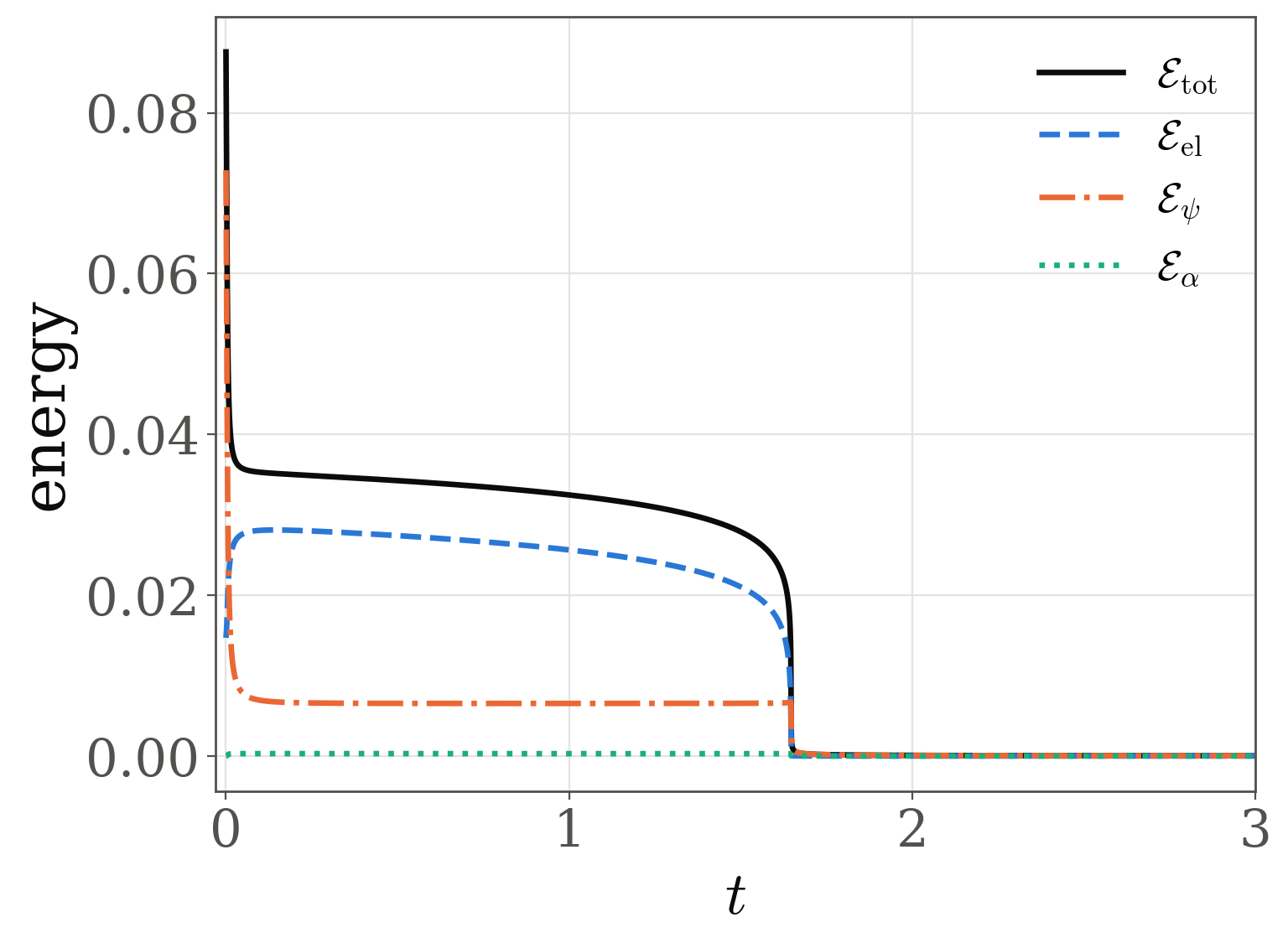}%
    }
    \\[0.35cm]
    \subfloat[]{%
        \label{fig:applied_stress_trapping-b}%
        \includegraphics[width=0.47\textwidth]{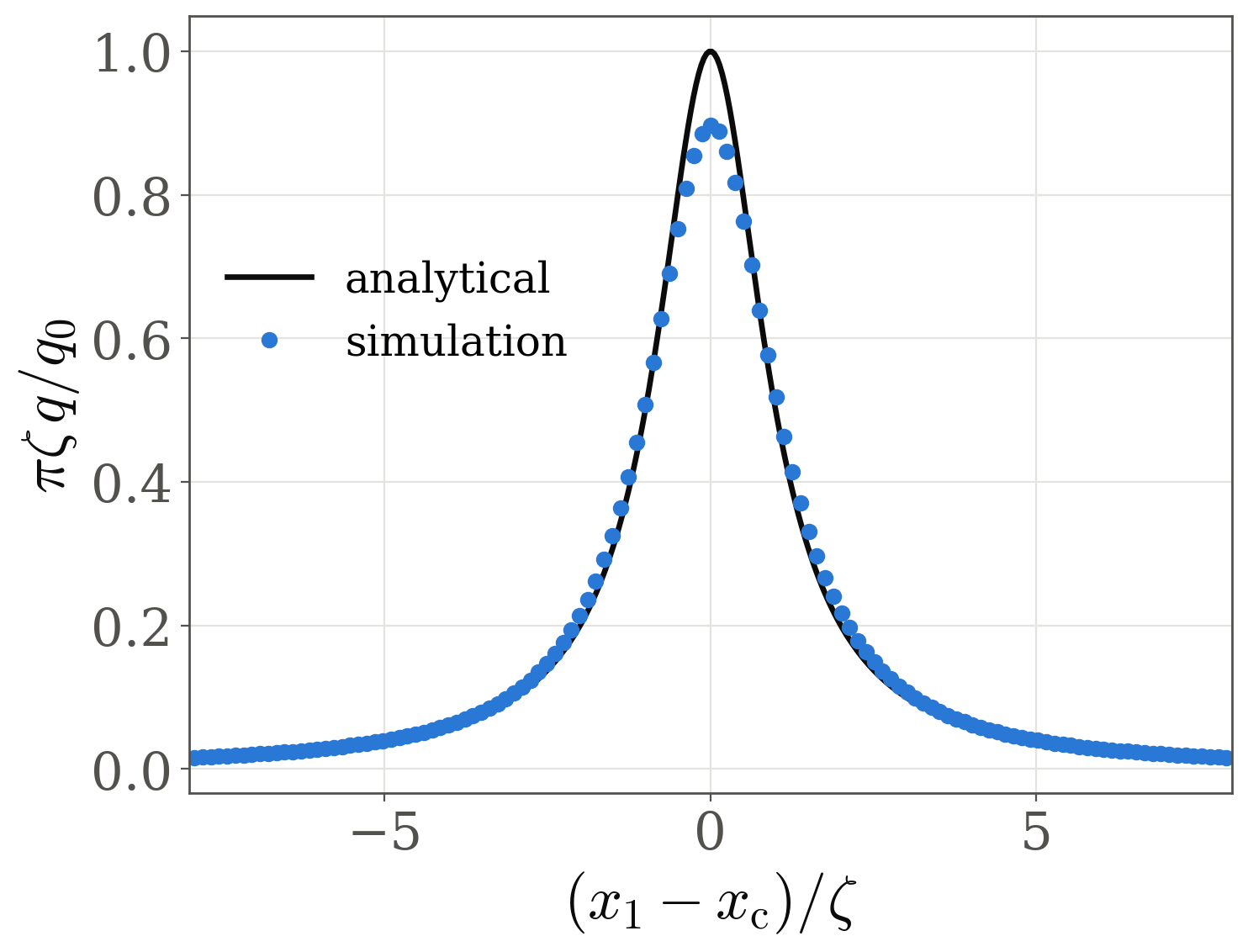}%
    }
    \caption{
    {
    Evolution of an opposite-sign dislocation pair at $T=0$.
    \textbf{(a)}
    Signed density $q(x_1,t)$ at several times.
    By $t=0.05$, the initial Gaussian cores have sharpened
    to nearly the Peierls--Nabarro width; they then glide
    toward each other and annihilate at $x_1=0$.
    \textbf{(b)}
    Elastic, core, lattice, and total energies along the
    evolution.
    \textbf{(c)}
    Profile of one core at $t=0.75$, centered at its
    maximum $x_{\rm c}$, compared with the
    Peierls--Nabarro core \eqref{eq:pn_core}.
    }
    }
    \label{fig:applied_stress_trapping}
\end{figure*}

\begin{figure*}[htb!]
    \centering
    \subfloat[]{%
        \label{fig:regime_II_stress-a}%
        \includegraphics[width=0.47\textwidth]{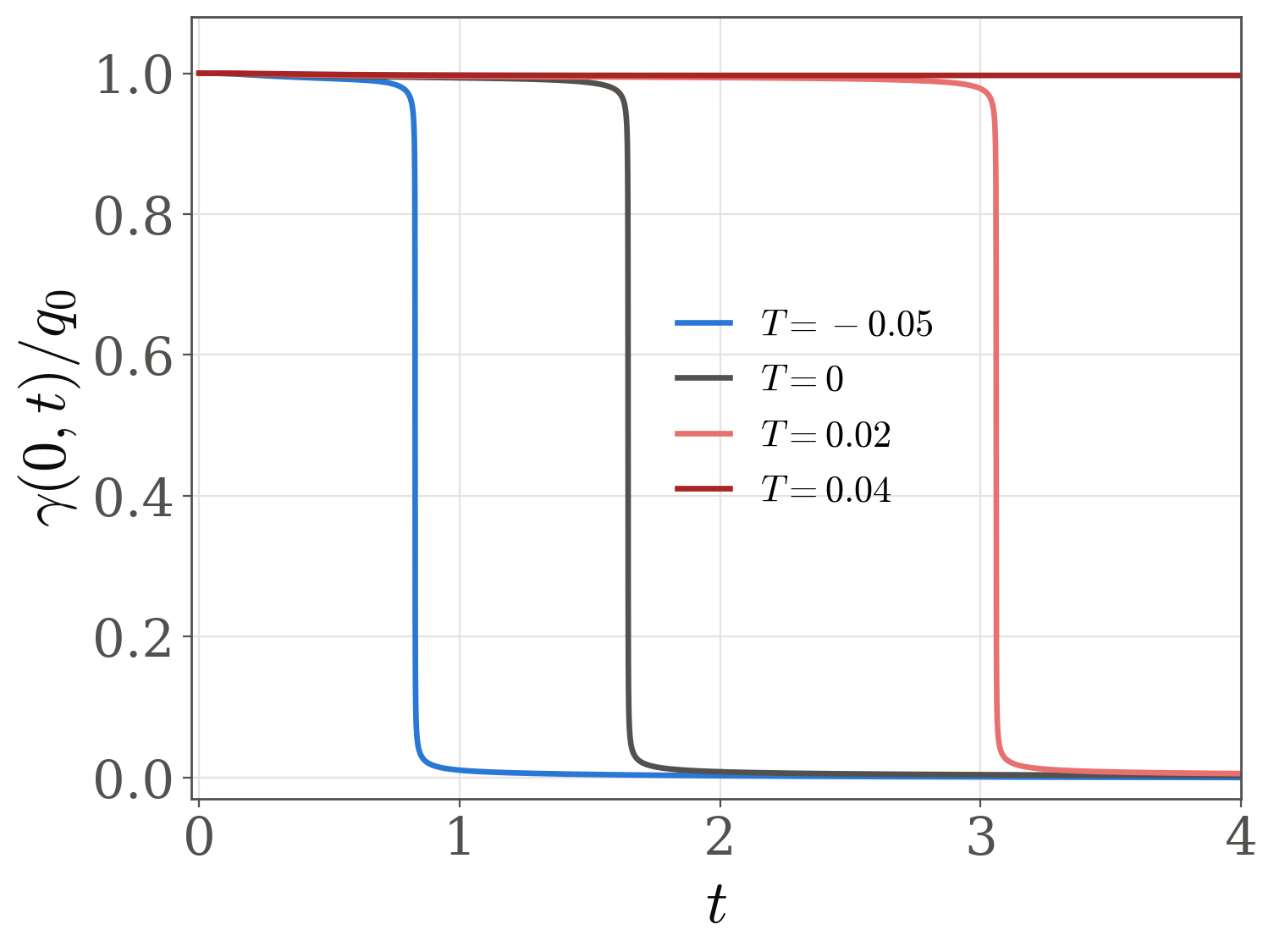}%
    }
    \hfill
    \subfloat[]{%
        \label{fig:regime_II_stress-b}%
        \includegraphics[width=0.47\textwidth]{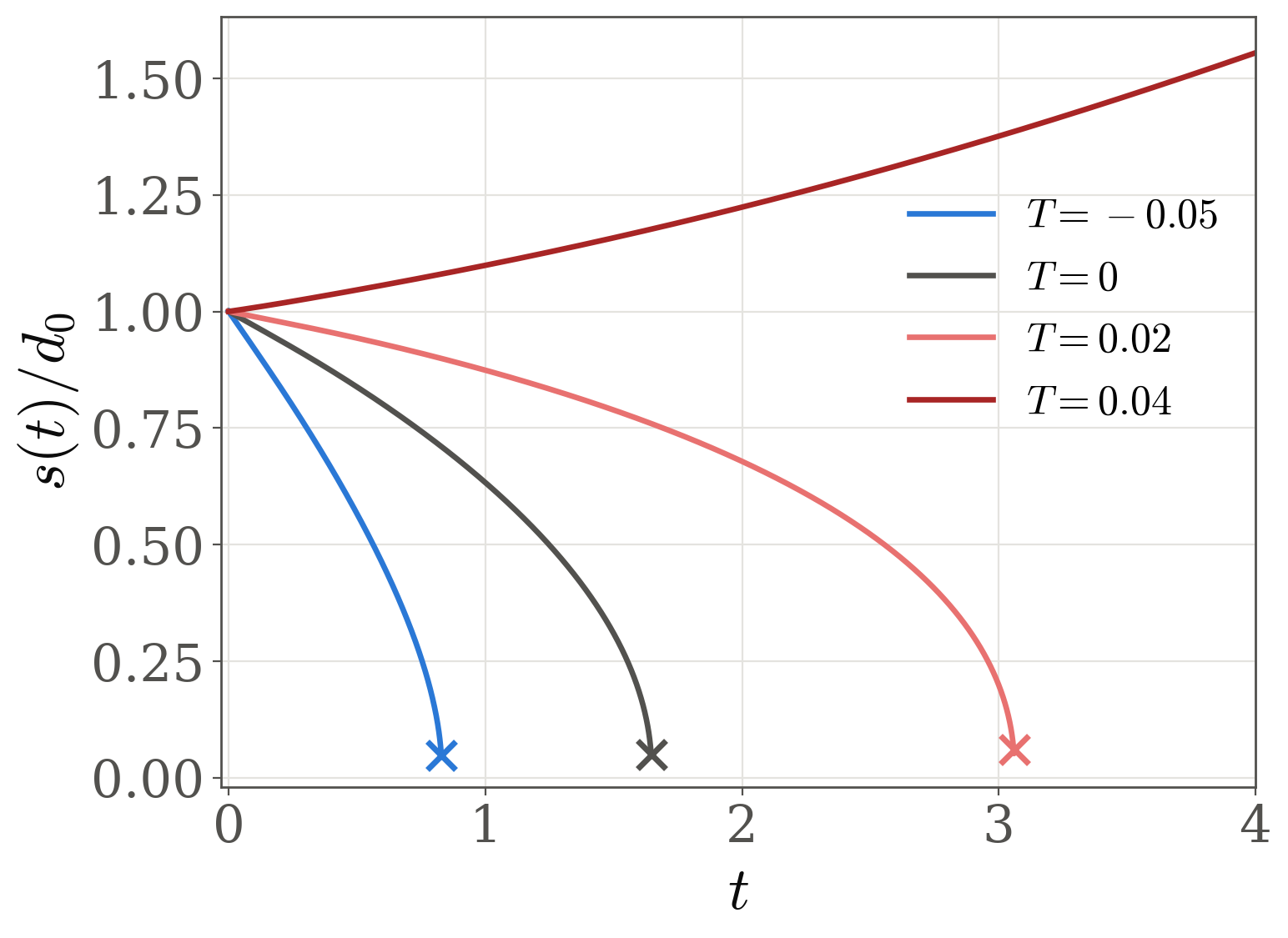}%
    }
    \caption{
    {
    The dislocation pair under a uniform applied stress $T$.
    \textbf{(a)}
    Midpoint slip $\gamma(0,t)$.
    \textbf{(b)}
    Centroid separation $s(t)$, up to annihilation
    (crosses, where $\gamma(0,t)=q_0/2$).
    Negative $T$ speeds up annihilation, small positive
    $T$ delays it, and $T>T_{\rm c}\approx0.031$
    separates the pair.
    }
    }
    \label{fig:regime_II_stress}
\end{figure*}

Thus, the
lattice energy keeps the cores narrow, so they glide as
compact Peierls--Nabarro cores and interact as point
dislocations.
An applied stress then either assists annihilation or,
above $T_{\rm c}$, separates the pair.

\begin{remark}[Spreading of a single core]
\label{rem:single_core}
When the elastic interaction dominates the lattice
energy, a dislocation core spreads instead of staying
compact and localized.
Consider a single dislocation at zero applied stress,
with $q(x_1,0)=q_0G_\varepsilon(x_1)$, $K=1$, and
$\kappa_h=0.10$, and with $q_0$, $b_{\rm slip}$, $m$,
$\varepsilon$, and $A_\psi$ as in
\eqref{eq:opposite_pair_numerical_parameters}, so that
$\Lambda=40$ and $\Gamma_\alpha=160$.
Since $q\geq0$ on $\mathbb R$, \eqref{eq:pair_velocities}
and \eqref{eq:general_reduced_pair_dynamics} hold on all
of $\mathbb R$.
The core spreads rapidly
(Fig.~\ref{fig:opposite_pair_numerical}).
Its half-width $w:=q_0/(\pi\max q)$, which equals $\zeta$
for the Peierls--Nabarro core \eqref{eq:pn_core}, grows
from $0.8\,\varepsilon$ to $6.7\,\varepsilon$ by $t=0.1$
and to about $7.6\,\varepsilon$ by $t=1$, where it has nearly stopped changing.
Since $\zeta/\varepsilon=\Lambda/(2\pi)\approx6.4$, the
steady core is about $20\%$ wider than the
Peierls--Nabarro core (dashed in Fig.~\ref{fig:opposite_pair_numerical}), because the core energy also
favors spreading.
We solve on $[-4,4]$ with grid spacing
$10^{-3}=\varepsilon/25$; doubling the grid spacing or the
domain size changes $w$ by less than $0.1\%$.
\end{remark}

\begin{figure*}[htb!]
    \centering
    \includegraphics[width=0.5\textwidth]{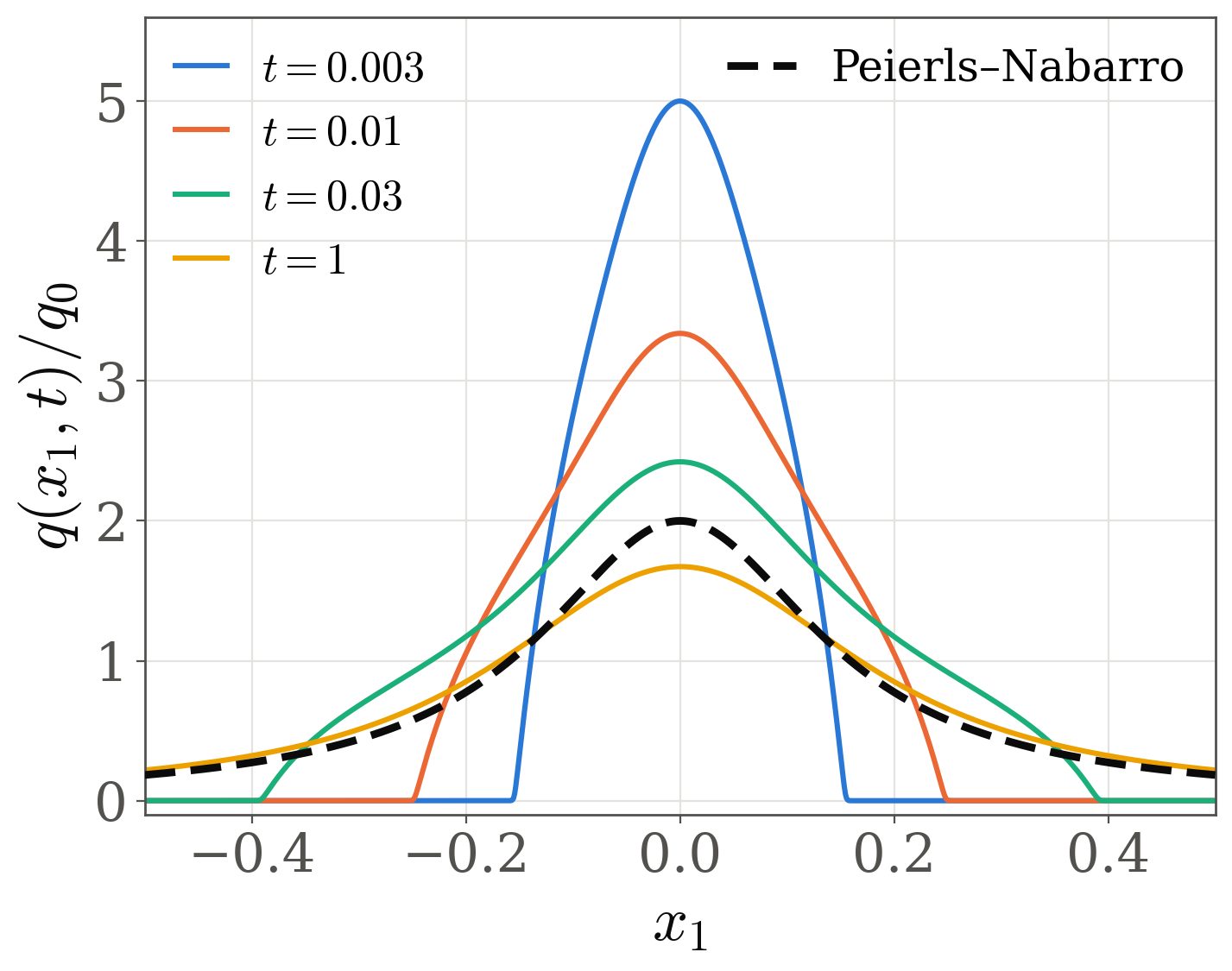}
    \caption{
    {
    Spreading of a single dislocation core when the elastic interaction dominates (Remark~\ref{rem:single_core}): signed density $q(x_1,t)$ at several times. The dashed curve is the Peierls--Nabarro core \eqref{eq:pn_core}.
    }
    }
    \label{fig:opposite_pair_numerical}
\end{figure*}

\subsection{Relation to Core and Lattice Energetics}
\label{subsec:comparison_defect_core_lattice}

The behaviors described above can be placed in relation to existing knowledge of defect, core, and lattice energetics in dislocation theory.

When the defect and lattice energies are retained, the present theory yields a
more general thermodynamic driving force. As shown in
\eqref{eq:alpha_force_general_new}, the force on \(\bfalpha\) contains the
effective-stress contribution involving
\(-\parderiv{\Psi}{\bfbeta_p}\), together with the
defect contribution involving
\(-\curl\mathbold P_{\alpha}^{\mathrm d}\).
The former is
generated by the nonconvex lattice energy, while the latter is generated by
the defect energy \(W_\alpha\). These terms are absent from stress-only
Peach--Koehler closures, e.g.,
\cite{Sethna2006,Sethna2007,Sethna2008,Xia2015,Monavari2018}.
However, related energetic contributions are present in other dislocation models.
Defect energies and their associated back-stress,
diffusion-stress, line-tension, gradient, and curvature contributions have
appeared in statistical and thermodynamic continuum dislocation theories
\cite{GromaCsikorZaiser2003,Zaiser2015,Hochrainer2016}.

Phase-field dislocation theories have likewise incorporated Peierls-type,
generalized stacking-fault, and multiwell lattice energies for
crystallographic slip, including atomistically informed energy landscapes
that represent lattice resistance and barriers to dislocation motion
\cite{Koslowski2002,Hunter2011,Peng2020}.
Peierls and Frenkel--Kontorova descriptions similarly represent lattice
energetics through nonconvex slip potentials and show that
lattice discreteness and nonlinearity can strongly influence dislocation
kinetics and stability under applied stress
\cite{Rosakis2001,Rosakis2013}. 
In Section~\ref{subsec:applied_stress_trapping}, for example, the balance of the elastic and lattice
energies selects nearly the Peierls--Nabarro core \eqref{eq:pn_core}, as in
these models.

}
\section{Concluding Remarks}
\label{sec:conclusion}

We have developed a transport-constrained variational gradient-flow formulation for continuum dislocation evolution.
The central idea is that the admissible rates for the gradient flow must be generated by dislocation transport.
In the formulation developed here, the plastic distortion is reconstructed from the dislocation density tensor and the compatible slip, with their evolution restricted to the transport directions obtained from geometric conservation principles.
The corresponding transport velocities are selected by minimizing a constrained Onsager functional, rather than being introduced independently as a closure such as the Peach--Koehler force.
The resulting theory has a formal gradient-flow structure compatible with the geometric dislocation conservation law; it provides the dissipative structure of a gradient flow while enforcing the geometric aspects of dislocation motion.

A key consequence of the gradient flow approach is that the dislocation driving force contains contributions beyond those that arise from elasticity, in contrast to the classical Peach--Koehler force on a dislocation.
When the lattice energy and core energy are neglected, the classical Peach--Koehler line force is recovered for a straight dislocation line.

We apply the model to predict the evolution of
two opposite-sign dislocations on a single glide
layer, where it reduces to one evolution equation for
the slip, which we solve numerically.
The lattice energy keeps the cores at
the Peierls--Nabarro width, and they glide toward each
other and annihilate.
An applied shear stress either assists annihilation or,
above a threshold set by the elastic attraction at the
initial separation, separates the pair.
The total energy drops rapidly as the cores form and as
they annihilate, and slowly as they glide toward each
other.

An important unexplored direction is the role of temperature.
The kinetics in this paper are athermal: the quadratic dissipation gives a linear force--velocity relation.
Unresolved activation barriers could be included through temperature-dependent, nonquadratic dissipation~\cite{KocksArgonAshby1975,MielkePeletierRenger2014}, which retains a generalized gradient structure.
Proposition~\ref{prop:transport_gradient_flow_identity} establishes energy decay for time-independent loading, but not convergence to an energy minimum, and the transport constraint may admit additional stationary states; for example, a dislocation-free state remains dislocation-free.
A rigorous understanding of the role of thermal effects and the interplay with nonconvexity is an important direction for the future.

\section*{}

\paragraph*{Software and Data Availability.}

The code developed for this work and the associated data are available at
\url{https://github.com/PariaBehnoud/transport-constrained-gradient-flow-dislocations}.

\paragraph*{Acknowledgments.}

We acknowledge John Clayton, Abigail Hunter, David Kinderlehrer, Jaroslaw Knap, Phoebus Rosakis, and Pradeep Sharma for insightful discussions; 
ARL (W911NF-25-2-0136) and ARO (MURI W911NF-25-2-0164) for financial support; 
and NSF-ACCESS (MTH250040) for computing resources provided by the Pittsburgh Supercomputing Center.


\makeatletter
\renewcommand*{\thesection}{\Alph{section}}
\renewcommand*{\theequation}{\thesection.\arabic{equation}}
\renewcommand*{\thesubsection}{\thesection.\arabic{subsection}}
\renewcommand*{\p@subsection}{}
\renewcommand*{\thesubsubsection}{\thesubsection.\arabic{subsubsection}}
\renewcommand*{\p@subsubsection}{}
\makeatother

\newcommand{\etalchar}[1]{$^{#1}$}

\end{document}

%% file: preamble.tex
\usepackage{isomath}
\usepackage{amsmath,amsthm}
\usepackage{amsbsy}
\usepackage{amssymb}
\usepackage{amscd}
\usepackage{amsfonts}
\usepackage{stmaryrd}
\usepackage{siunitx}
\usepackage{euscript}
\usepackage[utf8]{inputenc}
\usepackage[T1]{fontenc}
\usepackage{newtxtext} 
\everymath{\displaystyle}
\usepackage{exscale}
\usepackage{microtype}
\usepackage{hyperref}
\usepackage{xcolor}
\usepackage{placeins}
\usepackage{soul}
\usepackage{graphicx}
\usepackage{tikz}
\usetikzlibrary{arrows.meta}
\usetikzlibrary{calc}
\usepackage{boxedminipage}
\usepackage{calc}
\usepackage[dvipsnames]{xcolor}
\graphicspath{ {media/} }
\usepackage[caption=false,justification=raggedright]{subfig}

\usepackage{setspace}
\usepackage{enumitem}
\setitemize{noitemsep,topsep=0pt,parsep=0pt,partopsep=0pt}
\setenumerate{noitemsep,topsep=0pt,parsep=0pt,partopsep=0pt}
\setdescription{noitemsep,topsep=0pt,parsep=0pt,partopsep=0pt}

\usepackage{soul} 
\usepackage[normalem]{ulem}

\usepackage{orcidlink}
\usepackage{siunitx}
\usepackage[small]{titlesec}

\titlespacing*{\section}{0pt}{12pt plus 4pt minus 2pt}{2pt plus 2pt minus 2pt}
\titlespacing*{\subsection}{0pt}{12pt plus 4pt minus 2pt}{2pt plus 2pt minus 2pt}
\titlespacing*{\subsubsection}{0pt}{12pt plus 4pt minus 2pt}{2pt plus 2pt minus 2pt}
\titlespacing*{\paragraph}{0pt}{12pt plus 4pt minus 2pt}{2pt plus 2pt minus 2pt}

\makeatletter
    \renewcommand*{\thesection}{\arabic{section}}
    \renewcommand*{\thesubsection}{\thesection.\Alph{subsection}}
    \renewcommand*{\p@subsection}{}
    \renewcommand*{\thesubsubsection}{\thesubsection.\arabic{subsubsection}}
    \renewcommand*{\p@subsubsection}{}
\makeatother

\input{definitions}

%% file: definitions.tex
\usepackage{upgreek}

\newcommand{\eps}{{\varepsilon}}

\newtheorem{proposition}{Proposition}[section]

\theoremstyle{definition}

\AtEndEnvironment{definition}{\null\hfill\qedsymbol}
\newtheorem{remark}{Remark}[section]
\AtEndEnvironment{remark}{\null\hfill\qedsymbol}

\AtEndEnvironment{example}{\null\hfill\qedsymbol}

\AtEndEnvironment{assumption}{\null\hfill\qedsymbol}

\newcommand{\bfeta}{\mathbold {\eta}}

\newcommand{\bfsigma}{\mathbold {\sigma}}

\newcommand{\bfalpha}{\mathbold {\alpha}}
\newcommand{\bfomega}{\mathbold {\omega}}
\newcommand{\bfbeta}{\mathbold {\beta}}

\newcommand{\bfxi}{\mathbold {\xi}}

\newcommand{\bfzero}{\mathbf{0}}

\DeclareMathOperator{\divergence}{div}
\DeclareMathOperator{\curl}{curl}

\newcommand{\parderiv}[2]{\frac{\partial #1}{\partial #2}}
\newcommand{\dm}{\ \mathrm{d}}

\newcommand{\bfb}{{\mathbold b}}

\newcommand{\bfu}{{\mathbold u}}

\newcommand{\bfx}{{\mathbold x}}

\newcommand{\bfA}{{\mathbold A}}

\newcommand{\bfM}{{\mathbold M}}

\newcommand{\bfR}{{\mathbold R}}

\newcommand{\bfV}{{\mathbold V}}

%% file: media/jmps_section7_schematic.tex
\begin{figure}[htb!]
\centering
\begin{tikzpicture}[
    scale=0.9,
    >=Stealth,
    font=\small,
    line cap=round,
    line join=round,
    initial/.style={black, line width=1.15pt},
    later/.style={black!60, line width=1.05pt, dashed}
]
\def\halfsep{1.8}  
\def\eps{0.96}     
\def\amp{1.32}     
\def\c{0.6}        
\def\e{0.7}        
\def\ampl{1.81}    
\def\gs{1.4}       

\begin{scope}[shift={(0,7.05)}]
    \node[anchor=west, font=\bfseries] at (-6.45,1.86) {(a)};
    \draw[thin] (-5.25,-1.20) rectangle (5.25,1.22);

    \fill[cyan!8] (-5.25,-0.24) rectangle (5.25,0.24);
    \draw[gray!65, dashed] (-5.25,-0.24)--(5.25,-0.24);
    \draw[gray!65, dashed] (-5.25,0.24)--(5.25,0.24);
    \draw[gray!55, dashed] (-5.25,0)--(5.25,0);

    \fill[red!35, opacity=0.26] (-\halfsep,0) ellipse (0.90 and 0.34);
    \fill[red!55, opacity=0.33] (-\halfsep,0) ellipse (0.43 and 0.21);
    \fill[blue!35, opacity=0.26] (\halfsep,0) ellipse (0.90 and 0.34);
    \fill[blue!55, opacity=0.33] (\halfsep,0) ellipse (0.43 and 0.21);

    \draw[very thick] (-\halfsep,0)--(-\halfsep,0.60);
    \draw[very thick] ({-\halfsep-0.16},0)--({-\halfsep+0.16},0);
    \draw[very thick] (\halfsep,0)--(\halfsep,-0.60);
    \draw[very thick] ({\halfsep-0.16},0)--({\halfsep+0.16},0);
    \node[above] at (-\halfsep,0.72) {$+q_0\boldsymbol e_1$};
    \node[above] at ( \halfsep,0.72) {$-q_0\boldsymbol e_1$};

    \draw[<->, thin] (4.65,-0.24)--(4.65,0.24);
    \node[above] at (4.65,0.33) {$h$};
    \node[anchor=west] at (5.33,0) {$x_2=0$};
    \node[anchor=west] at (-4.90,0.78) {glide layer};
    \draw[->, thin] (-3.72,0.67)--(-3.45,0.22);

    \draw[->] (-5.10,1.50)--(-3.98,1.50)
        node[right] {$x_1$};
    \draw[->] (-5.10,1.50)--(-5.10,2.10)
        node[above] {$x_2$};
    \draw (3.15,1.66) circle (0.125);
    \fill (3.15,1.66) circle (0.035);
    \node[anchor=west] at (3.38,1.66) {$x_3$ (out of plane)};
\end{scope}

\begin{scope}[shift={(0,2.55)}]
    \node[anchor=west, font=\bfseries] at (-6.45,1.95) {(b)};
    \draw[->, thin] (-5.35,0)--(5.45,0)
        node[below] {$x_1$};
    \draw[->, thin] (-5.15,-1.75)--(-5.15,1.95)
        node[above] {$q$};
    \node[anchor=east] at (-5.23,0) {$0$};

    \draw[initial, samples=190, smooth, domain=-5.1:5.1, variable=\x]
        plot (\x,{\amp*(exp(-((\x+\halfsep)^2)/(2*\eps^2))
                      -exp(-((\x-\halfsep)^2)/(2*\eps^2)))});
    \draw[later, samples=190, smooth, domain=-5.1:5.1, variable=\x]
        plot (\x,{\ampl*(exp(-((\x+\c)^2)/(2*\e^2))
                       -exp(-((\x-\c)^2)/(2*\e^2)))});

    \draw[thin] (-\halfsep,0.07)--(-\halfsep,-0.07);
    \draw[thin] ( \halfsep,0.07)--( \halfsep,-0.07);
    \node[below] at (-\halfsep,-0.08) {$-d_0/2$};
    \node[above] at ( \halfsep,0.08) {$d_0/2$};

    \draw[<->, thin] (-\halfsep,1.52)--({-\halfsep+\eps},1.52)
        node[midway, above=1pt] {$\varepsilon$};

    \draw[->, thick, red!75!black] (-3.05,-0.95)--(-1.65,-0.95);
    \draw[->, thick, blue!75!black] (3.05,0.95)--(1.65,0.95);

    \node[anchor=east] at (-2.75,1.05) {$q(x_1,0)$};
    \node[black!60, anchor=west] at (1.10,1.62) {$q(x_1,t)$};
    \draw[->, thin, black!60] (1.05,1.55)--(-0.62,1.42);
\end{scope}

\begin{scope}[shift={(0,-1.45)}]
    \node[anchor=west, font=\bfseries] at (-6.45,1.85) {(c)};
    \draw[->, thin] (-5.35,0)--(5.45,0)
        node[below] {$x_1$};
    \draw[->, thin] (-5.15,-0.25)--(-5.15,1.85)
        node[above] {$\gamma$};
    \node[anchor=east] at (-5.23,0) {$0$};

    \draw[initial, samples=190, smooth, domain=-5.1:5.1, variable=\x]
        plot (\x,{\gs*(1/(1+exp(min(9,-1.702*(\x+\halfsep)/\eps)))
                      -1/(1+exp(min(9,-1.702*(\x-\halfsep)/\eps))))});
    \draw[later, samples=190, smooth, domain=-5.1:5.1, variable=\x]
        plot (\x,{\gs*(1/(1+exp(min(9,-1.702*(\x+\c)/\e)))
                      -1/(1+exp(min(9,-1.702*(\x-\c)/\e))))});

    \draw[gray!70, densely dashed] (0,0)--(0,1.40);
    \fill[black!60] (0,0.87) circle (0.045);
    \node[black!60, anchor=west] at (0.95,1.62) {$\gamma(0,t)$};
    \draw[->, thin, black!60] (0.92,1.58)--(0.08,0.93);
    \node[below] at (0,-0.04) {$0$};

    \node[anchor=east] at (-2.55,1.20) {$\gamma(x_1,0)$};
    \node[black!60, anchor=west] at (2.95,0.78) {$\gamma(x_1,t)$};
    \draw[->, thin, black!60] (2.92,0.74)--(1.18,0.30);
\end{scope}
\end{tikzpicture}
\caption{{Schematic of the opposite-sign edge-dislocation pair.
(a) The dislocations have Burgers vectors $\pm q_0\boldsymbol e_1$ and
line direction $\boldsymbol e_3$, on a thin glide layer of thickness $h$
centered at $x_2=0$.
(b) Signed density $q$: initial state with two Gaussian cores of standard
deviation $\varepsilon$ centered at $x_1=\mp d_0/2$ (solid), and a later
state (dashed). The velocity reverses direction with the sign of $q$; when the
net stress drives the cores together, positive and negative Burgers content
move toward $x_1=0$ (arrows) and cancel there.
(c) Slip $\gamma$, with $q=\partial_{x_1}\gamma$; the slip $\gamma(0,t)$
equals the Burgers content on either side of $x_1=0$ and decreases as the
pair annihilates.
All dimensions and profiles are schematic.}}
\label{fig:opposite_pair_geometry}
\end{figure}